\documentclass[%
 reprint,
 amsmath,amssymb,
 aps,
]{revtex4-2}

\usepackage{graphicx}
\usepackage{dcolumn}
\usepackage{bm}
\usepackage{multirow}

\usepackage{hyperref}
\hypersetup{
 linktocpage = true,
     colorlinks,
    citecolor=blue,
    filecolor=black,
    linkcolor=blue,
    urlcolor=blue
}

\begin{document}

\preprint{APS/123-QED}


\title{Persistent Homology of $\mathbb{Z}_2$ Gauge Theories}

\author{Dan Sehayek}
\affiliation{%
 Department of Physics and Astronomy, University of Waterloo, Ontario, N2L 3G1, Canada}
 \affiliation{
 Perimeter Institute for Theoretical Physics, Waterloo, Ontario N2L 2Y5, Canada
 }
 
\author{Roger G. Melko}
\affiliation{%
 Department of Physics and Astronomy, University of Waterloo, Ontario, N2L 3G1, Canada}
 \affiliation{
 Perimeter Institute for Theoretical Physics, Waterloo, Ontario N2L 2Y5, Canada
}

\date{\today}

\begin{abstract}

Topologically ordered phases of matter display a number of unique characteristics, including 
ground states that can be interpreted as patterns of closed strings in the case of general $\mathbb{Z}_2$ string liquids. 
In this paper, we consider the problem of detecting and distinguishing closed strings in Ising spin configurations
sampled from the classical $\mathbb{Z}_2$ gauge theory.
We address this using the framework of persistent homology, which computes the size and frequency of general loop structures in spin configurations via the formation of geometric complexes.
Implemented numerically on finite-size lattices, we show that the first Betti number of the Vietoris-Rips complexes achieves a high density at low temperatures in the $\mathbb{Z}_2$ gauge theory.  In addition, it displays a clear signal at the finite-temperature deconfinement transition of the three-dimensional theory. 
We argue that persistent homology should be capable of interpreting prominent loop structures that occur in a variety of systems, making it a useful tool in theoretical and experimental searches for topological order.

\end{abstract}

\maketitle


\section{\label{sec:intro}Introduction} 

Topological order is a key concept in modern condensed matter physics, lying beyond Landau's paradigm of an order parameter defined by symmetry breaking.
The most well-known example of topological order is the class of general quantum string liquids \cite{Wen1,Wen2,Kitaev}, which refers to systems where the ground state wavefunction is an equal superposition over all closed string configurations. Such an ordering gives rise to exotic properties such as emergent fractionalized excitations and topological degeneracy on closed manifolds, which besides their theoretical intrigue may hold technological importance in the field of topological quantum computing \cite{Nayak}.  Therefore, a major effort is underway to search for and identify topologically ordered phases in materials \cite{Balents2010}, devices \cite{Chen2021,Satzinger}, and synthetic quantum matter \cite{RydbergSL}.

The challenge of detecting topological order usually translates into examining the system's configuration space. Due to the lack of an order parameter, this can involve using tools such as the topological entanglement entropy \cite{TOPOEE1,TOPOEE2}. In situations where system configurations are represented by data, such as in computer simulation studies
or in projective measurements of a quantum device, such tools can be prohibitively computationally expensive.  This motivates the search for
interpretable features which are inherently sensitive to topological structures in data, while remaining tractable on large finite-size
lattice systems of interest to condensed matter and quantum information physics.

Topological data analysis -- the field that focuses on assigning geometry to general point clouds -- provides a number of powerful tools for the extraction of topological features from data.
Chief among them, {\it persistent homology} is a tool for computing the significance and frequency of topological features, such as loops and connected clusters, via the computation of a sequence of geometric complexes \cite{TDA1,TDA2,TDA3,TDA4}. Recently, the application of persistent homology to condensed matter systems has been explored. In Ref. \cite{PH1}, phase transitions in the mean-field XY model and classical $\phi^4$ model were detected by directly computing the persistent homology of the configuration space. A more recent perspective promotes persistent homology as an observable \cite{Sale}. Here, configurations are sampled from a physical model, and persistence diagrams of each of these configurations are computed. Such diagrams contain the significance and frequency of loop structures, defined by the geometric complex considered. In Refs.~\cite{PH2,PH3}, phase transitions in the XXZ model on a pyrochlore lattice, 2D XY model and 1D quantum models are detected by clustering configurations based on their persistence diagram using appropriate kernel methods. In Refs.~\cite{PH4,Sale}, persistence diagrams are instead transformed into persistence images, and classified using logistic regression and k-nearest neighbours methods. Ref.~\cite{Sale}, which studies various XY models, was the first to show that accurate estimates for critical exponents of the correlation length can be obtained from finite-scaling analysis of persistent homology observables.

In this paper, we demonstrate the utility of persistent homology on configurations where the feature of interest is closed loop configurations of Ising variable $\sigma^z_i = \pm 1$.
We consider samples of these configurations obtained from thermal states of the classical $\mathbb{Z}_2$ gauge theory in $D=2$ and 3 spatial dimensions \cite{Kogut,MCP},
\begin{equation}
    H=-K\sum_{\square}P_{\square}, \label{Ham}
\end{equation}
where the Ising degrees of freedom are placed on the edges of a square plaquette, $P_\square=\prod_{\ell\in\square}\sigma_\ell^z$ and the groundstate is satisfied 
when all $P_\square=1$ on the lattice.
We define the persistence of a loop structure via a sequence of Vietoris-Rips (VR) complexes, which can be easily calculated directly from Ising configurational data. We demonstrate the calculation of the frequency of topological features as measured by persistent homology in the $2D$ and $3D$ gauge theories. In the latter case, we find that the expectation value of the first Betti number of VR complexes at a fixed radius ($1/2$ the lattice spacing) can accurately detect the deconfinement transition directly from individual Ising configurations. This demonstrates that the full persistence measure is not necessarily needed to identify the phase transition. Our work illustrates persistent homology analysis as a tool for interpreting the types of loop structures that form in construction of the VR complex. 

\section{\label{sec:TDA}Topological analysis with homology}

Topology is concerned with global properties of shapes: properties that remain unchanged under smooth transformations that avoid cutting and pasting. Formally, such transformations are \textit{homeomorphisms}: bijective and continuous maps with a continuous inverse. If such a map exists between two objects, then the two objects are regarded as \textit{topologically equivalent}. Such equivalence classes can be defined more intuitively in terms of \textit{topological invariants}: properties that remain invariant under homeomorphisms. One example of this is the set of Betti numbers of a manifold $\mathcal{M}$, where the $n^\text{th}$ Betti number, denoted $b_n$, corresponds to the number of $n+1$ dimensional holes of $\mathcal{M}$. Formally, $b_n$ is given by the rank of the $n^\text{th}$ homotopy group of $\mathcal{M}$ (or the $n^\text{th}$ homology group). For a brief introduction to homotopy and homology, the reader is encouraged to read Appendix \ref{app:homotopy} and \ref{app:homology}.

\begin{figure}[t]
    \centering
    \includegraphics[width=0.5\textwidth]{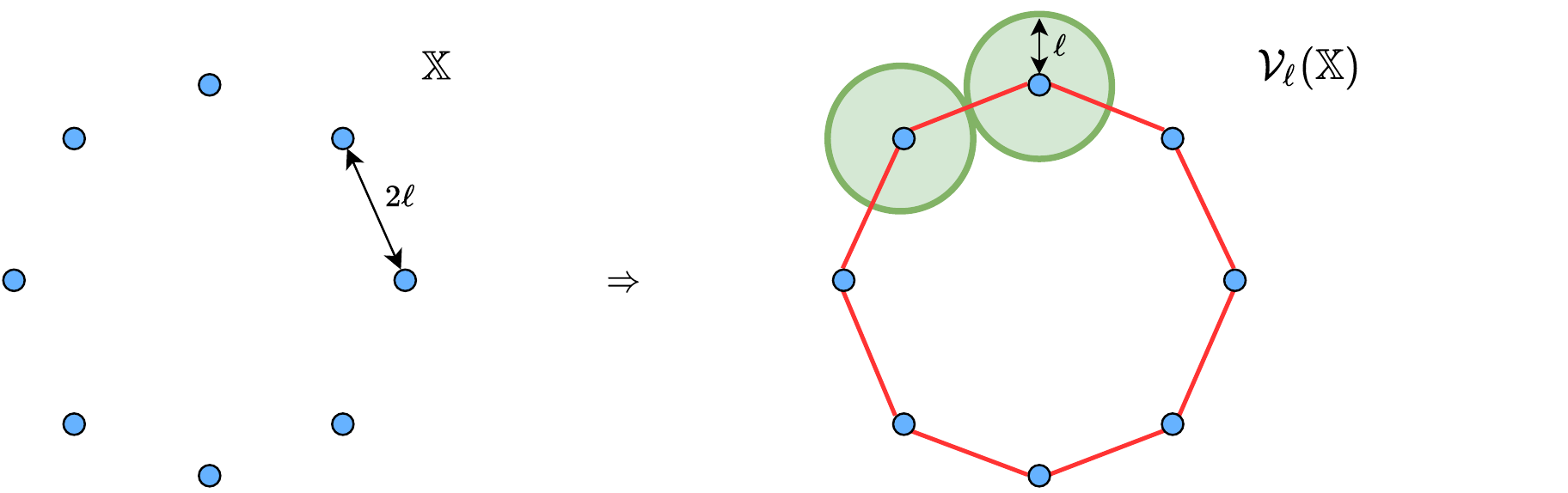}
    \caption{Point cloud $\mathbb{X}$ and associated VR complex $\mathcal{V}_\ell(\mathbb{X})$. At a radius of $r=\ell$, connections (1-simplices) form between all neighbouring points. The resulting complex is homeomorphic to a circle, and ultimately has a first Betti number of $b_1=1$. }
    \label{fig:2dvr}
\end{figure}

\subsection{\label{sec:vrcomplex}Vietoris-Rips complex}

The Vietoris-Rips (VR) complex is a tool for extending homology to point clouds \cite{TDA2}. As an example, we consider the set of points $\mathbb{X}$ shown in Figure \ref{fig:2dvr}. As a set of points, $\mathbb{X}$ has no topology ($b_n=0$ $\forall n$). However, it is clear that $\mathbb{X}$ has an underlying loop structure. To identify this loop structure, one can consider the $\ell$ VR complex of $\mathbb{X}$, denoted $\mathcal{V}_\ell(\mathbb{X})$. In this simple case, the $\mathcal{V}_\ell(\mathbb{X})$ is constructed by forming connections (1-simplices or lines) between neighbouring points (all points within a distance of $2\ell$ away from each other). The resulting graph, which is homeomorphic to a circle, clearly has a nontrivial first homology group given by $H_1(\mathcal{V}_\ell(\mathbb{X}))\cong\mathbb{Z}$. Ultimately, the VR complex construction has allowed us to assign a topology to $\mathbb{X}$, and the underlying loop structure has been successfully identified.

Generally, the $r$ VR complex $\mathcal{V}_{r}(\mathbb{X})$ of a point cloud $\mathbb{X}$ is constructed by forming an $n$-simplex (see Appendix \ref{app:simplices}) for every subset of $n+1$ data points, denoted $\left \{ x_0,...,x_n \right \}\subset{\mathbb{X}}$, satisfying $d(x_i,x_j)\leq{2r}$, where $d$ is the selected metric. For a $D$ dimensional point cloud, this can be visualized as expanding $D$-spheres of radius $r$ around each point. If the spheres of two points overlap, a 1-simplex (line) is formed using these points. If the spheres of three points all overlap with each other, the a 2-simplex (filled triangle) is formed using these points. Generally, the formation of higher simplices is important for identifying higher homologies. For example, the nontrivial second homology associated with a 2-sphere can only be successfully identified under the formation of 2-simplices.

\begin{figure}[t]
    \includegraphics[width=\columnwidth]{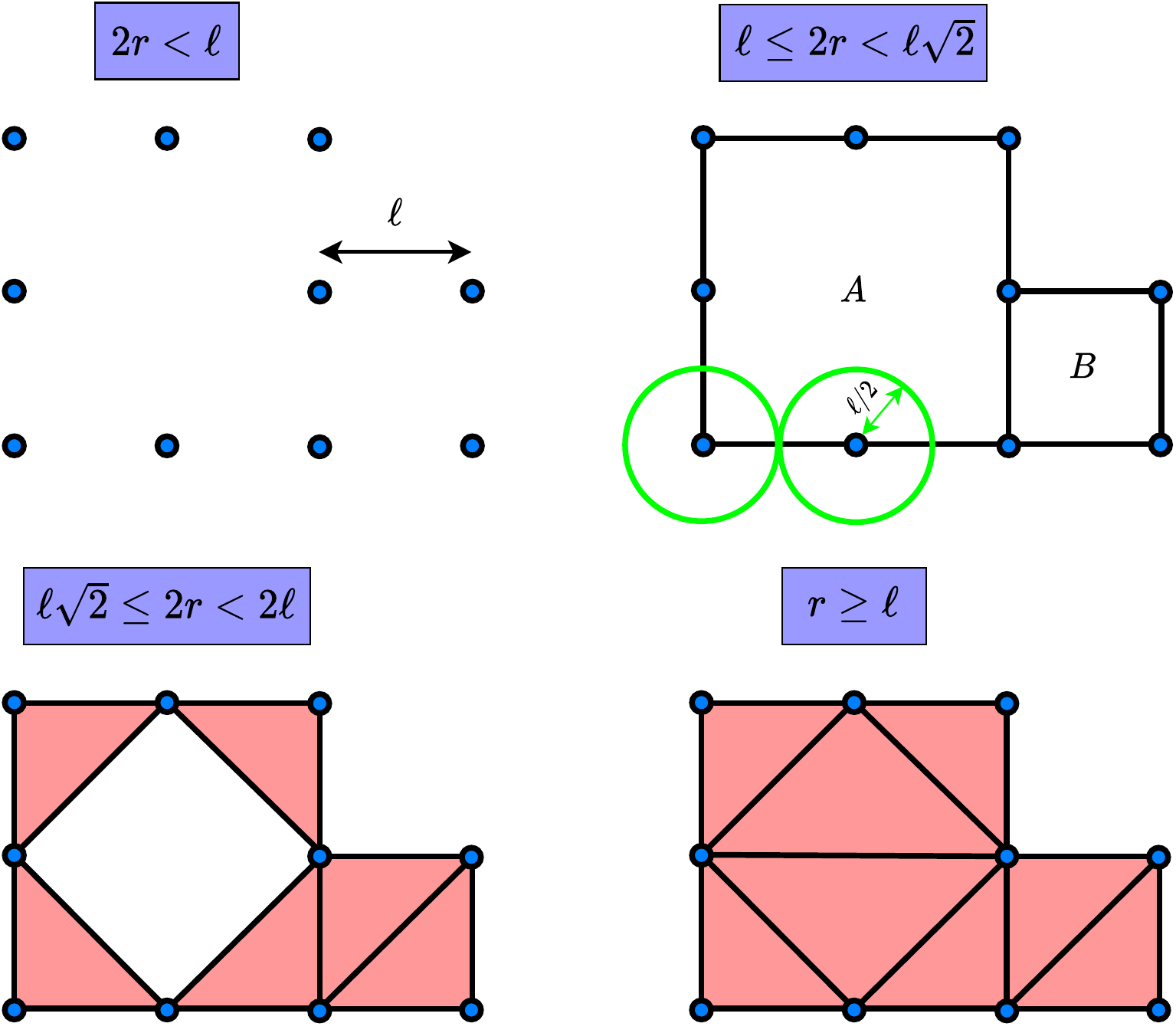}
    \caption{Illustration of persistent homology applied to a set of data points (blue dots). For any radius $2r<\ell$, no circles overlap, and hence, no connections form. At a radius of $2r=\ell$, all points within a distance of $\ell$ from each other form connections. The corresponding graph has 2 holes, labelled $A$ and $B$. Since these holes form exactly at $2r=\ell$, they are said to be \textit{born} at $2r=\ell$. At a radius of $2r=\ell\sqrt{2}$, diagonal connections of length $\ell\sqrt{2}$ form, and the resulting simplices (triangles) are filled in (pink). Since hole $B$ is no longer present, it is said to have \textit{died} at $2r=\ell\sqrt{2}$. Similarly, at a radius of $r=\ell$, one final connection forms, and two more simplices are filled in. Since hole $A$ is no longer present, it is said to have \textit{died} at $r=\ell$. To measure the significance of each hole, one can compute the \textit{persistence} as death minus birth. These are given by $p_A=2\ell-\ell=\ell$ and $p_B=\ell\sqrt{2}-\ell\approx{0.4\ell}$ for holes $A$ and $B$, respectively. Since hole $A$ has a greater persistence than hole $B$, it is considered to be a more significant topological feature in the data.}
    \label{fig:phomology}
\end{figure}

\subsection{\label{sec:phomology}Persistent homology on point clouds}

At this point, several questions arise. How does one know what radius to choose? And how does one measure the significance of these loop features? In regards to the latter question, one can imagine a scenario in which tinier loop structures exist on the edges of the circle considered in Figure \ref{fig:2dvr}. We consider such loop structures, which could be formed for example due to noise in a dataset, as being of less interest. To address these questions, one can compute the VR complexes $\mathcal{V}_r(\mathbb{X})$ and the resulting homology groups $H_n(\mathcal{V}_r(\mathbb{X}))$ for many different values of the radius $r$. The resulting sequence of VR complexes, with each element in the sequence corresponding to the formation of new simplices, is referred to as the \textit{filtration} \cite{TDA2}. If a given loop structure forms at a radius $r_b$ in the filtration and dies at a radius $r_d$ in the filtration, then the significance (persistence) is measured by the difference in these radii: $p=r_d-r_b$.

To illustrate this procedure, we consider the example shown in Figure \ref{fig:phomology}. As connections form and simplices are filled in, holes in the corresponding graph form and disappear. In this particular example, two holes, labelled $A$ and $B$, form at a radius of $2r=\ell$, where $\ell$ is the minimal distance between points. Holes $A$ and $B$ then disappear at a radius of $2r=\ell\sqrt{2}$ and $r=\ell$, due to the filling of simplices. Generally, the radius at which a hole forms and disappears is referred to as the \textit{birth} and \textit{death} of that hole, respectively. The significance of the corresponding loop structure in the data can then be measured as death minus birth, which is referred to as the \textit{persistence} of the hole. In this case, the persistence of holes $A$ and $B$ are $p_A=2\ell-\ell=\ell$ and $p_B=\ell\sqrt{2}-\ell\approx{0.4\ell}$, respectively. Hence, $A$ is regarded as a more significant topological feature in the data, since $p_A>p_B$. Persistence can be understood in terms of stability \cite{Stability}: topological features with greater persistence are more stable against perturbations in the data. Ultimately, this procedure allows one to detect and measure the significance of all loop structures in the data, and is referred to as \textit{persistent homology}.

\begin{figure}[t]
    \centering
    \includegraphics[width=0.47\textwidth]{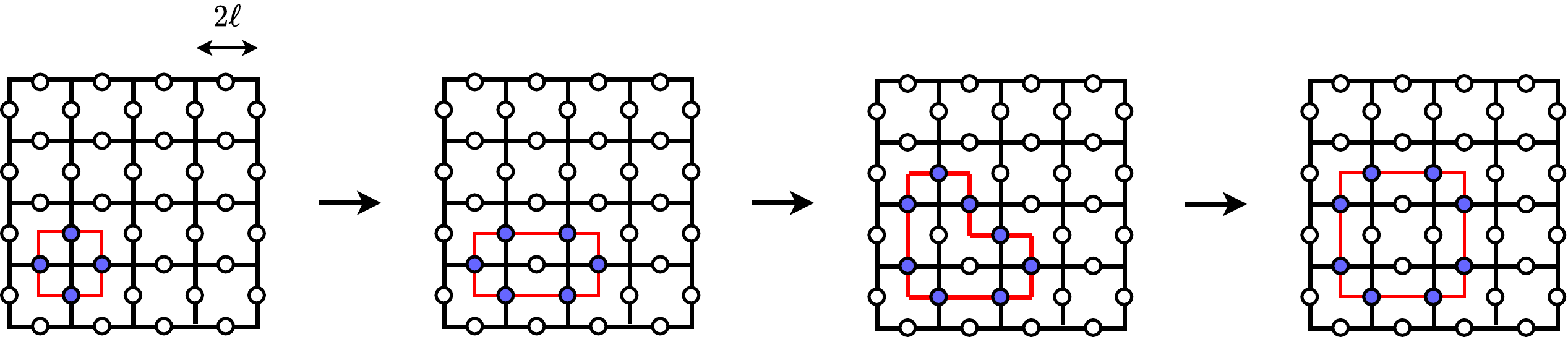}
    \caption{Examples of closed strings in ground state configurations of the $\mathbb{Z}_2$ gauge theory, where blue dots depict down spins. The application of a single gauge transformation to a configuration consisting solely of up spins generates a closed string of minimal size. Arrows depict steps in which single adjacent gauge transformations are applied, which stretches this loop to greater sizes. In the context of persistent homology, this corresponds to generating loops of greater persistence. The first configuration corresponds to an $H_1$ homology that is born at $r=\ell/\sqrt{2}$ and dies at $r=\ell$. The remaining configurations correspond to $H_1$ homologies that are born at $r=\ell$ and die at $r=\left \{ \ell\sqrt{2},\ell\sqrt{5/2},\ell\sqrt{9/2} \right \}$ respectively.}
    \label{fig:loops}
\end{figure}

\subsection{Homology for the $\mathbb{Z}_2$ gauge theory} \label{sec:HZ2GT}

The ground state wavefunction of a general spin-1/2 string liquid can be understood as a quantum superposition over all closed string configurations. In the case of the toric code \cite{Kitaev}, this holds for both the $\sigma^x$ basis and $\sigma^z$ basis. More precisely, if strings are formed over links of the direct (dual) lattice for which $\sigma^x=-1$ ($\sigma^z=-1$), then all strings in any ground state configuration will be closed. Since we are considering the classical $\mathbb{Z}_2$ gauge theory described in Eq. \eqref{Ham}, a similar procedure for identifying closed strings on the dual lattice applies.


Alternatively, these closed strings can be defined using the VR complex. If $\bm{\sigma}$ represents a configuration of spins, then we begin by defining $X_{\bm{\sigma}}$ as a graph consisting of a 0-simplex (point) at the location of each down spin. Closed strings can then be identified as nontrivial $H_1$ homologies of the corresponding VR complex at $r=\ell$: $\mathcal{V}_\ell(X_{\bm{\sigma}})$ where $2\ell$ is the lattice spacing. Ultimately, each closed string will contribute to the first Betti number:
\begin{equation}
    b_1(\bm{\sigma}):=\text{rank}(H_1(\mathcal{V}_\ell(X_{\bm{\sigma}}))),
\end{equation}
while the size of each closed string can be measured by the persistence. Namely, in the persistent homology computation of $X_{\bm{\sigma}}$, each closed string will correspond to an $H_1$ birth-death point with $r_b=\ell$, while the size of each closed string can be measured as $r_d-r_b$. We note that $b_1$ accounts for all closed strings except those of minimal size (those generated by applying a single gauge transformation to an initial configuration consisting of all up spins). Such loops correspond to $H_1$ birth-death points with $r_b=\ell/\sqrt{2}$ and $r_d=\ell$. Examples of closed strings are shown in Figure \ref{fig:loops} for the case of a $D=2$ square lattice.

\begin{figure}[t]
    \centering
    \includegraphics[width=0.4\textwidth]{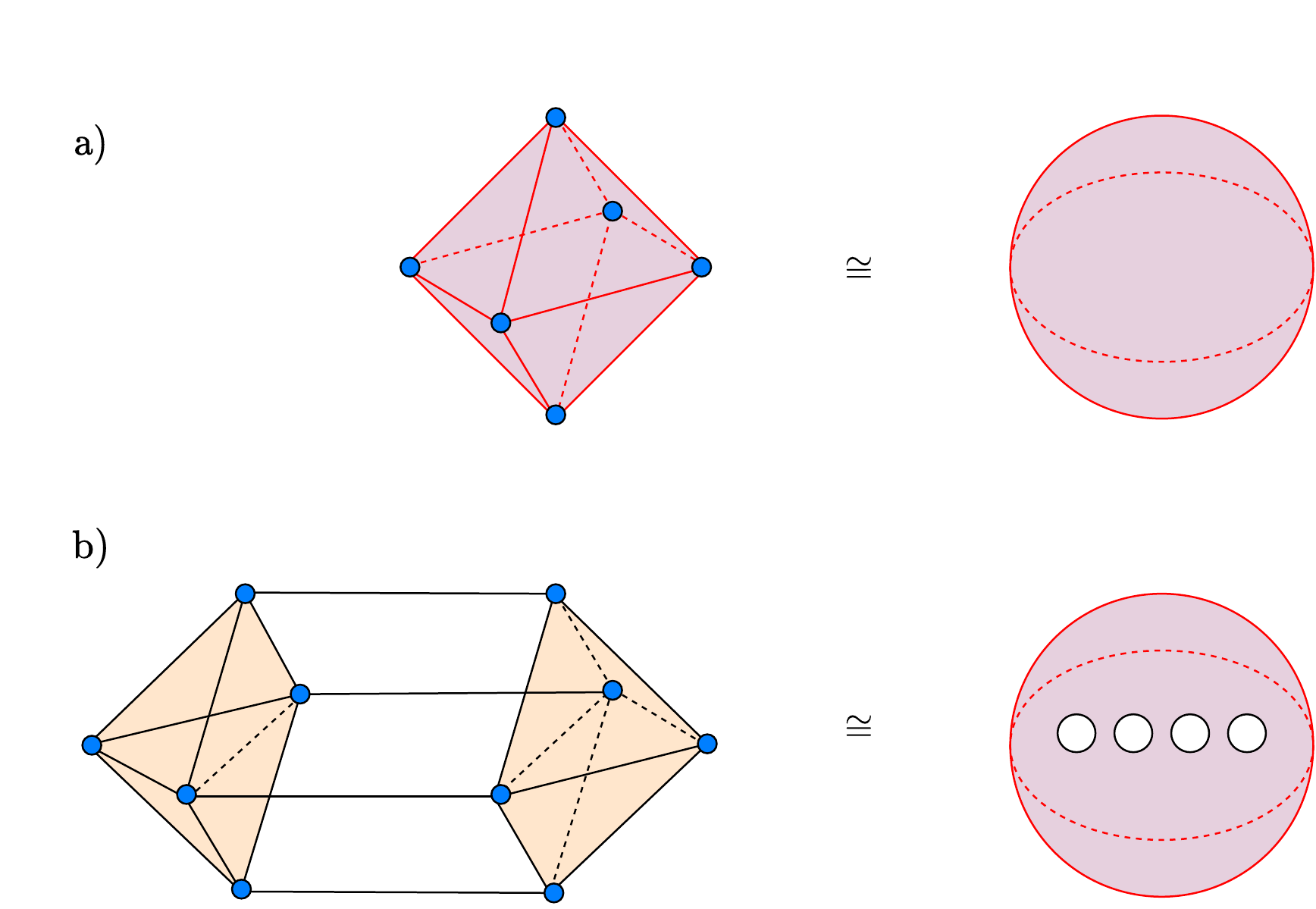}
    \caption{Top (bottom): Illustration of VR complex corresponding to $\bm{\sigma}'$ ($\bm{\sigma}''$), which is generated by applying one (two) gauge transformations to a configuration consisting solely of up spins. Blue points depict down spins. Red structures are hollow and beige structures are filled. The top VR complex of $\bm{\sigma}'$ holds generally for $\ell/\sqrt{2}\leq{r}<{\ell}$. The bottom VR complex of $\bm{\sigma}''$ holds generally for $\ell\leq{r}<\ell\sqrt{3/2}$. $\cong$ denotes homotopic equivalence.}
    \label{fig:h3d}
\end{figure}

What about ground state configurations of the classical $\mathbb{Z}_2$ gauge theory in $D=3$? In Figure \ref{fig:h3d}, we show the VR complexes of graphs $X_{\bm{\sigma}'}$ and $X_{\bm{\sigma}''}$, where $\bm{\sigma}'$ ($\bm{\sigma}''$) are generated by applying one (two) gauge transformations to an initial configuration consisting solely of up spins. The complex shown in Figure \ref{fig:h3d}a for $\bm{\sigma}'$ holds generally for $\ell/\sqrt{2}\leq{r}<{\ell}$. Such a complex, which is homeomorphic to a hollow sphere, possesses second Betti number $b_2=1$.

The complex shown in Figure \ref{fig:h3d}b for $\bm{\sigma}''$ holds generally for $\ell\leq{r}<\ell\sqrt{3/2}$. Such a complex is homeomorphic to a sphere with 4 holes. This in turn is homeomorphic to a plane with 3 holes: simply imagine expanding one hole. Hence, the corresponding first Betti number is $b_1=3$.

Applying a string of adjacent gauge transformations can then be understood as ``gluing'' planes with holes. Generally, if $\bm{\sigma}^{n}$ is generated by applying $n$ adjacent gauge transformations to a configuration consisting of all up spins, then the number of holes is given by $b_1(\bm{\sigma}^n)=3n$. On the other hand, generating vison defects through spin flips can be understood as eliminating connections (1-simplices) in the complexes considered above. Hence, vison defects will generally reduce the number of holes, and one expects $\langle{b_1}\rangle$ to decrease with increasing $T$. 

\section{\label{sec:det}Results}

To generate configuration data for the classical $\mathbb{Z}_2$ gauge theory in arbitrary dimensions, we employ a standard classical Monte Carlo method, consisting of cluster updates ($\mathbb{Z}_2$ gauge transformations) and local updates (spin flips) to sample the state at finite temperature according to the Metropolis algorithm. 
Then, persistent homology calculations are done on this configurational data using the Python Ripser package \cite{Ripser}. To reduce computational runtime, we set the \textit{threshold} parameter (the radius $r$ at which to stop the persistent homology computation) to 2 for the $3D$ case. It is additionally possible to restrict the computation to lower homology groups.
Furthermore, to account for the periodic boundary conditions, we use the following modified metric:
\begin{equation}
    d(\bm{x},\bm{y})=\sqrt{\sum_{\alpha=1}^D\min\left [ y_\alpha-x_\alpha,\left ( x_\alpha-\ell_1\right )+\left ( \ell_2-y_\alpha \right ) \right ]^2}
\end{equation}
where $y_\alpha>x_\alpha$ and $\ell_2>\ell_1$ are the locations of the boundaries. In our case, $\ell_1=0$ and $\ell_2=L$, and any data point $\bm{x}$ will correspond to the location of a down spin on the lattice. 

Generally, the $n^\text{th}$ Betti number $b_n$ of the $\mathcal{V}_r$ complex of a Monte Carlo configuration $\bm{\sigma}$ will be given by the number of $H_n$ birth-death points in the persistent homology computation with birth value $r_b\leq{r}$ and $r_d>r$. In the remaining sections, we will compute the expectation value of $\langle{b_1}\rangle$ averaged over an ensemble of Monte Carlo sampled configurations for various temperatures $T$.

\subsection{Two dimensional $\mathbb{Z}_2$ gauge theory}

\begin{figure}
    \centering
    \includegraphics[width=0.5\textwidth]{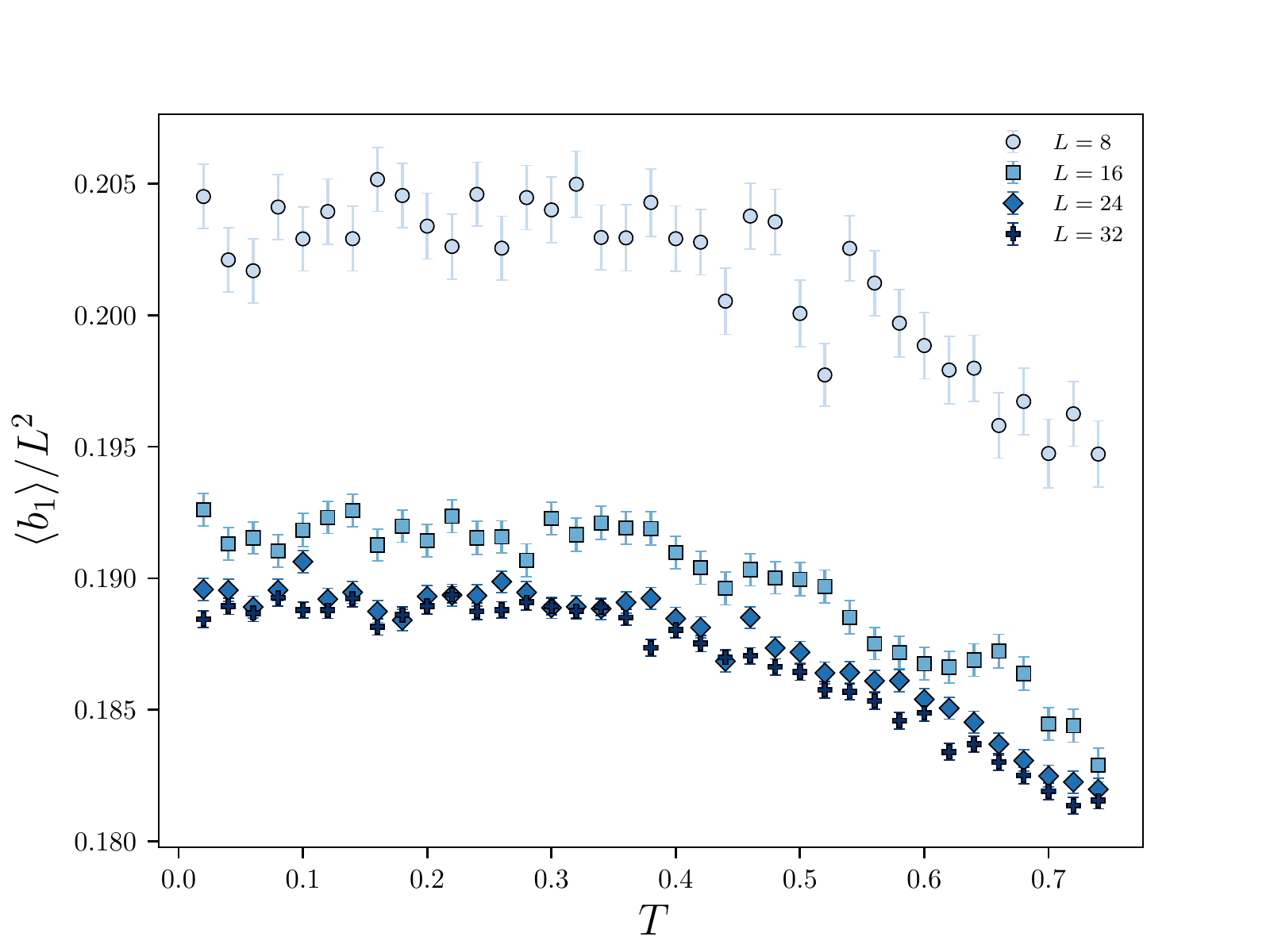}
    \caption{Average first Betti number of $r=\ell$ VR complexes of configurations in the $\mathbb{Z}_2$ gauge theory with $D=2$. $\langle{b_1}\rangle$ is averaged over 2000 samples for each temperature. Standard errors bars defined as $99\%$ Gaussian confidence intervals.}
    \label{fig:2d}
\end{figure}

We begin by examining the $D=2$ case. In Figure \ref{fig:2d}, average loop densities $\langle{b_1}\rangle/L^2$ are shown for various temperatures and lattice sizes $N=2\times L \times L$, with 2000 Monte Carlo samples used for each temperature. Generally, there exists $N=D\times{L}^D$ degrees of freedom, since there exists $L^D$ vertices on the lattice, and $D$ unique links can be associated with each vertex. We find that the frequency of closed strings generally decreases with increasing $T$. This is expected: as the temperature increases, vison defects (open strings) generate and hence reduce the number of closed strings. 

Figure \ref{fig:2d} additionally reveals that smaller system sizes possess slightly larger loop densities. In other words, we find that $\langle{b_1}\rangle$ is not perfectly extensive. Recalling Figure \ref{fig:loops}, the size of a closed loop can be measured by the corresponding persistence value, which is given by $p=r_d-r_b$. In Table \ref{tab:T1}, we show the total frequency of loops with various death values and birth value $r_b=\ell$. Here, we observe that the the frequency of loops with death value $r_d=3\ell$ is larger for $L=8$, despite being the smallest system size. On the other hand, loops with death values $r_d>3\ell$ occur for all sizes but $L=8$. 
This is consistent with the deviation of $\langle{b_1}\rangle$ from perfect extensiveness. While their are no dynamics in a loop condensate, it would make sense that the frequency of loops of larger sizes is restricted by the size of the lattice. If larger loops are more likely to form for larger system sizes, then one would expect the average loop density to decrease. Such an argument is weaker when comparing two larger system sizes, since there is an overall limit to the size of loops that can form.

\begin{table}[t]
\begin{tabular}{c|cccc|cccc}
\hline \hline
\( r_d/\ell \) & $\sqrt{2}$ & $\sqrt{2.5}$ & $\sqrt{4.5}$ & $3$ & $\sqrt{10}$ & $\sqrt{12.5}$ & $\sqrt{13}$   \\
 \hline
$L=8$ & 124.73 & 151.30 & 57.47 & 59.55 & 0 & 0 & 0 \\  \hline
$L=16$ & 125.65 & 150.90 & 59.99 & 1.28 & 35 & 9 & 1 \\  \hline
$L=24$ & 125.63 & 149.28 & 60.16 & 2.20 & 164 & 59 & 3 \\  \hline
$L=32$ & 124.59 & 149.45 & 60.14 & 3.06 & 562 & 188 & 12 \\
\end{tabular}
\caption{Frequencies of $H_1$ birth-death points with birth value $r_b=\ell$ and various death values $r_d$. Based on 2000 Monte Carlo samples used to generate Figure \ref{fig:2d} with $T=0.02$. No vison defects are present. Frequencies for $2\leq{r_d}\leq{3}$ are divided by $L^2$ and rounded to 2 decimal places to illustrate minimal changes in density.}
\label{tab:T1}
\end{table}

This data confirms the general expectation that $\langle{b_1(T)}\rangle$ decreases with increasing temperature.  However, since the $2D$ gauge theory has no phase transition at any non-zero temperature,
the data in Figure \ref{fig:2d} shows no sharp features or non-analytic behavior.  One can ask 
the question whether this quantity behaves different in the presence of a phase transition.
To this end, we next consider the $3D$ case, which has a critical point at $T_c \approx 1.313$. In other words, we ask under the VR complex construction, can the first Betti number be used to detect the deconfinement transition between the topologically ordered and disordered regimes?

\subsection{Three dimensional $\mathbb{Z}_2$ gauge theory}

\begin{table}[t]
\begin{tabular}{c|ccc|cccc}
\hline \hline
 & & $r_b=\ell$ & & & $r_b=\ell/\sqrt{2}$ & &    \\
 \hline
\( r_d/\ell \)& $\sqrt{3/2}$ & $\sqrt{2}$ & $\sqrt{5/2}$ & $\sqrt{3/2}$ & $\sqrt{2}$ & $2$ & $>2$    \\
 \hline
$L=6$ & 774.52 & 8.76 & 0.05 & 97.65 & 7.09 & 13.88 & 0 \\  \hline
$L=10$ & 779.81 & 8.93 & 0.10 & 99.25 & 10.06 & 0 & 3.00 \\  \hline
$L=14$ & 778.77 & 8.85 & 0.09 & 98.73 & 11.27 & 0 & 1.09 \\  \hline
$L=18$ & 778.82 & 8.87 & 0.09 & 98.81 & 11.72 & 0 & 0.51 \\
\end{tabular}
\caption{Total frequencies of $H_1$ birth-death points over 1000 Monte Carlo samples (divided by $L^3$). Based on 1000 Monte Carlo samples used to generate Figure \ref{fig:3d} with $T=0.7$. No vison defects are present. Frequencies are divided by $L^3$ and rounded to 2 decimal places.}
\label{tab:T2}
\end{table}

\begin{figure}
    \centering
    \includegraphics[width=0.5\textwidth]{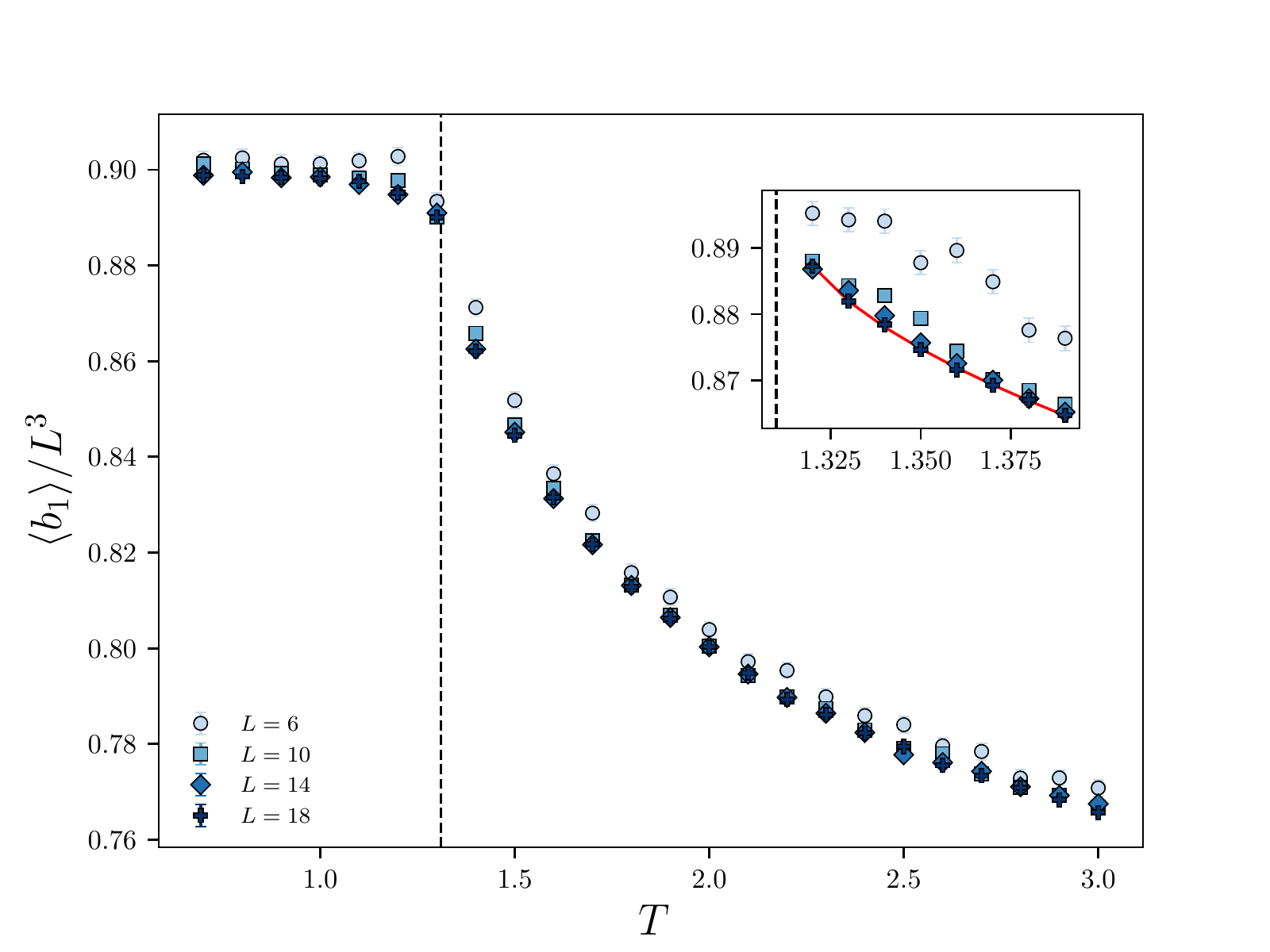}
    \caption{Average first Betti number of $r=\ell$ VR complexes of configurations in the $\mathbb{Z}_2$ gauge theory with $D=3$. $\langle{b_1}\rangle$ is averaged over 1000 samples for each temperature. Dashed line indicates theoretical value of critical temperature. Location of the phase transition is correctly identified by a change in the concavity of $\langle{b_1}\rangle$. Statistical error bars are not visible.}
    \label{fig:3dall}
\end{figure}

To examine the $D=3$ case, we use our Monte Carlo simulation of Eq.~\eqref{Ham} on a cubic lattice to produce 1000 samples at each temperature for various lattice sizes.  Figure \ref{fig:3dall} shows the value of $\langle{b_1}\rangle$ for various linear system sizes $L$ as a function of temperature. Similar to the $D=2$ case, we find that the first Betti number increases with decreasing temperature, as expected in the case of decreasing vison defects, and that smaller system sizes have slightly larger loop densities. Once again, we see that from Table \ref{tab:T2} that is due to the presence of larger loops in larger system sizes. Namely, loops with birth value $r_b=\ell/\sqrt{2}$ and death value $r_d=2\ell$ occur only in $L=6$, while loops with the same birth value and larger death values only occur in the larger system sizes. Among these larger system sizes, we see that the frequencies of loops with $r_d>2\ell$ decreases with increasing system size, which again agrees with the idea that larger loops are more prominent in larger system sizes, hence reducing the overall frequency of loops.

The temperature range in Figure \ref{fig:3dall} includes the known value of the critical temperature in the $3D$ model, at $T_c \approx 1.314$.  It is clear from the data that $\langle{b_1}\rangle$ has a sharp feature at $T_c$, indicating that this quantity is sensitive to critical fluctuations that are manifest in the loop structures present in the spin configurations.
Above the critical temperature, we see a particularly rapid rise in $\langle{b_1}\rangle$ as the transition is approached. We find that this can be accurately fit to $\langle{b_1(T)}\rangle\sim(T-T_c)^{\phi}$, where $\phi\approx{0.544}$ is an estimate obtained from the largest system size considered in Figure \ref{fig:3dall}. We note that while the behavior at the critical point is consistent with a divergence, a standard finite-size scaling analysis \cite{Cardy} using $\phi$ and the correlation length critical exponent for the $3D$ Ising model does not produce a higher-quality data collapse than already observed in Figure \ref{fig:3dall}.

\begin{figure}
    \centering
    \includegraphics[width=0.5\textwidth]{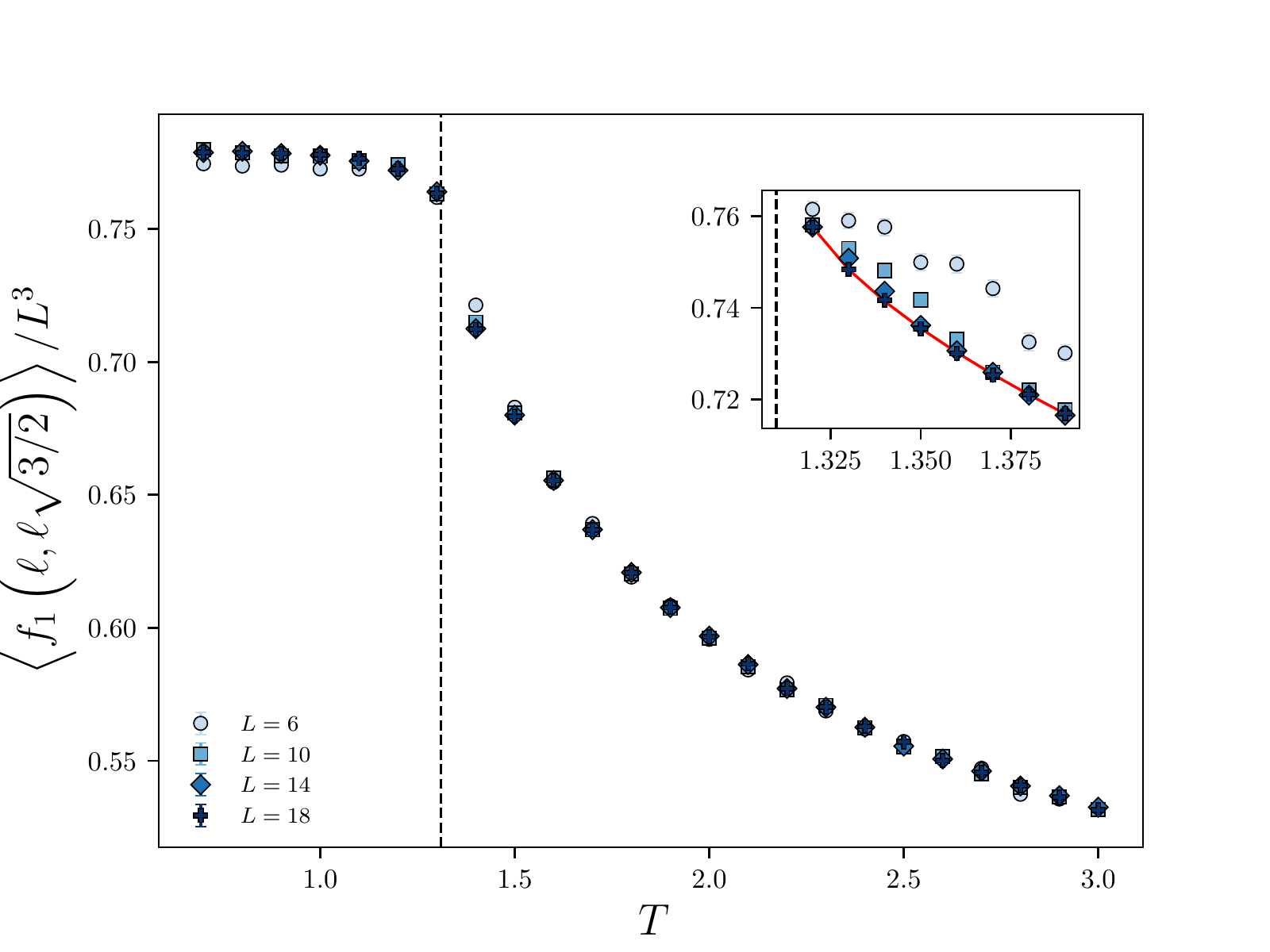}
    \caption{Average frequency of $H_1$ homologies with birth and death values $r_b=\ell$ and $r_d=\ell\sqrt{3/2}$. Based on the same Monte Carlo configurations used in Figure \ref{fig:3dall}. }
    \label{fig:3d}
\end{figure}

Finally, we note that the full Betti number (all loops with birth values $r_b\leq\ell$ and death values $r_d>\ell$) is not required to detect this transition. As an example, Figure \ref{fig:3d} shows the frequency of loops with birth value $r_b=\ell$ and death value $r_d=\ell\sqrt{3/2}$. As before, we find that the deconfinement transition is detected by a sharp feature in the average frequency. Note the exact structure of these loops was discussed in Section \ref{sec:HZ2GT}. Once again, if we assume that the expectation value takes the form $(T-T_c)^\phi$ for $T>T_c$, then we can obtain a value of $\phi\approx{0.582}$ from the largest system size considered. In the case of the second Betti number $b_2$, we find that all non-trivial second homologies with birth value $r_b=\ell$ have the same death value $r_d=\ell\sqrt{3/2}$. Namely, we find that that $b_2=f_2(\ell,\ell\sqrt{3/2})$. Once again, $\left \langle b_2 \right \rangle$ displays a sharp feature at $T_c$ and can be accurately fit to $\left \langle b_2\right \rangle\sim(T-T_c)^\phi$ where $\phi\approx{0.644}$.

\begin{figure}[t]
    \centering
    \includegraphics[width=0.4\textwidth]{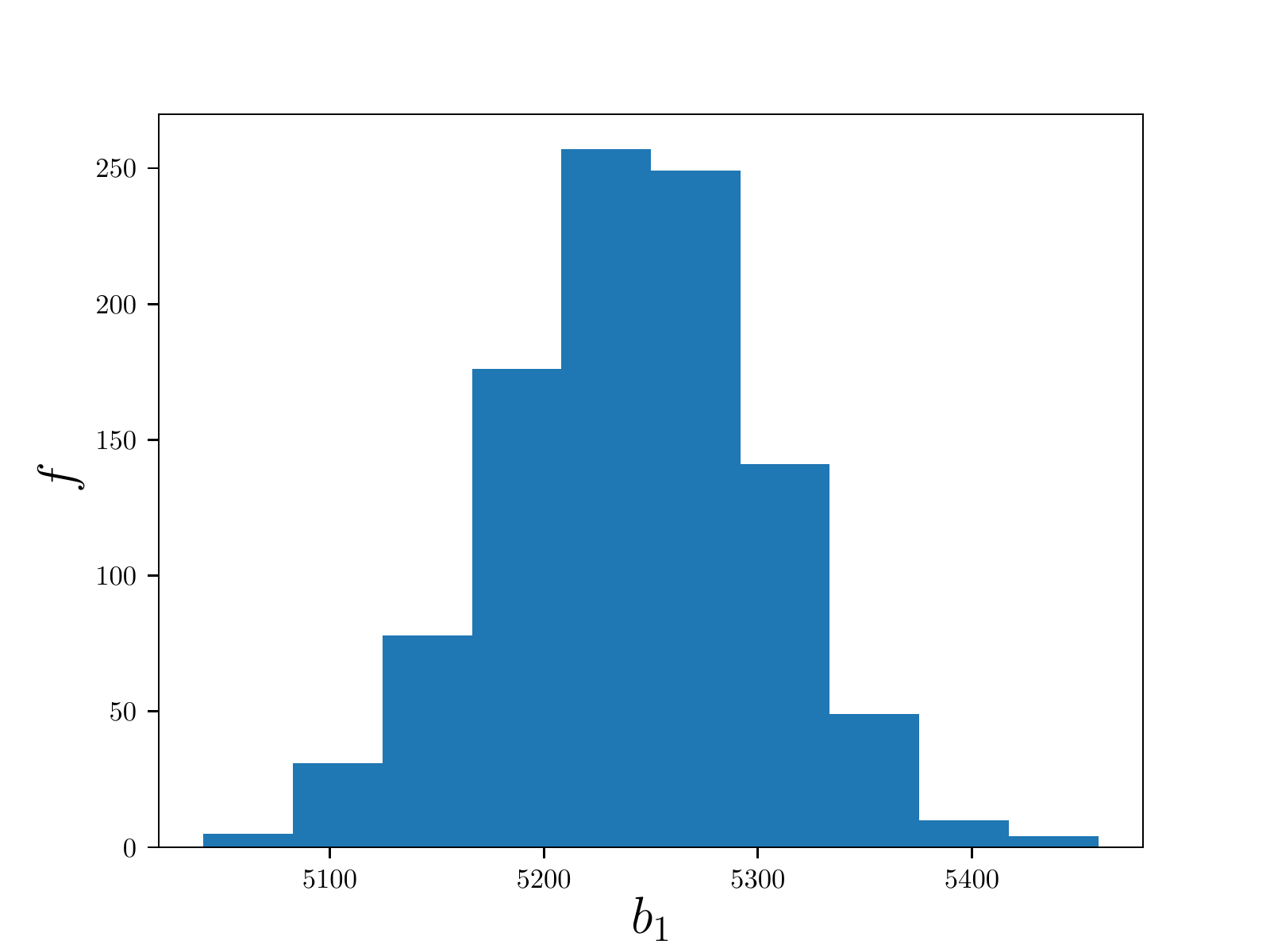}
    \caption{Histogram of number of Monte Carlo samples $f$ for $b_1$ obtained under the VR complex construction of samples from the $3D$ $\mathbb{Z}_2$ gauge theory, with $L=18$ and $T=0.7$. Based on the same Monte Carlo samples used to generate Figure \ref{fig:3dall}. }
    \label{fig:betti}
\end{figure}

\section{\label{sec:discussion}Discussion}

In this paper, we explore the idea of assigning topologies to configurations in the pure $\mathbb{Z}_2$ gauge theory using the Vietoris-Rips (VR) complex. Following construction of the VR complex, the first Betti number $\langle b_1 \rangle$ can be computed via simplicial homology. We obtain thermal configurational data for the $\mathbb{Z}_2$ gauge theory in dimensions $D=2,3$ using Monte Carlo simulations. We find that the expectation value of this first Betti number is largest at lowest temperatures, where the configurations are generally expected to satisfy the constraints of the Hamiltonian model, indicating that the measurement is sensitive to loop (or closed string) structures in the configurational data.  As the temperature increases, vison defects (open strings) are generated, which reduce the number of closed strings, and correspondingly reduce the value of $\langle b_1 \rangle$. In the $3D$ model, this quantity is successful in identifying the previously-known critical point, and is consistent with a functional form $(T-T_c)^{\phi}$ with an exponent $\phi\approx{0.544}$.

Compared to machine learning perspectives for detecting general phase transitions, our approach has several advantages. To begin with, this method does not possess any learning parameters nor any hyperparameters, as is the case in neural network models. Hence, there is no parameter complexity, and no required tuning or hyperparameter grid search. In terms of the minimum number of samples required to correctly identify the phase transition, we expect that the corresponding complexity decays as $1/L^3$, since the first Betti number is shown to be extensive to a very good approximation.

Another well-established method for the detection of topological order in the $\mathbb{Z}_2$ gauge theory include computation of the Wegner-Wilson loop and clustering metrics. While the Wegner-Wilson loop is certainly easier to compute, the construction of such a gauge-invariant observable requires knowledge of the gauge transformation defining the local symmetry. In regards to clustering metrics, it was shown in \cite{DM1} that the diffusion map algorithm can successfully identify topological sectors defined by the sign of the average Wegner-Wilson loop. However, the metric used to compute the kernel matrix once again required knowledge of the gauge transformation, and additionally posed a heavy bottleneck. In contrast, computation of the first Betti number of the VR complex requires no previous knowledge of the system, and is completely interpretable. Namely, a general persistent homology analysis will yield the exact birth and death values and frequencies for each identified loop structure in any given configuration, with no {\it a priori} knowledge of the gauge structure underlying the model.

A next logical step would be to consider quantum models with $\mathbb{Z}_2$ topological order. One example would be the toric code with external field terms $h_X\sum\sigma^x$ and $h_Z\sum\sigma^z$, which can be mapped to the classical $3D$ $\mathbb{Z}_2$ gauge theory with a uniform field \cite{MCP}. Such a model possesses two distinct deconfinement transitions characterized by the condensation of $e$ particles (for $h_X\gg{1}$) and $m$ particles (for $h_Z\gg{1}$). In the 2+1D membrane picture and the $\sigma^x$ basis, $h_X$ can be viewed as a parameter for the surface tension of membranes (since $\sum\sigma^x$ is a string tension term) and $h_Z$ can be viewed as a parameter for the frequency of holes in the membranes (since $\sum\sigma^z$ is a generator for pairs of $e$ particles) \cite{Somoza}. In this case, we expect that the frequency and persistence values for $H_1$ and $H_2$ homologies may serve as useful tools for detecting and distinguishing the deconfinement transitions.

It would of additional interest to explore the behaviour of these Betti numbers in phase transitions not characterized by anyon condensation. One example would be the Haldane phase with hidden $\mathbb{Z}_2\times\mathbb{Z}_2$ symmetry breaking \cite{Tasaki}. The prominent structure of this nontrivial symmetry-protected topological (SPT) phase is closed strings of +1 and -1 spins, as defined by the string order parameter of den Nijs and Rommelse \cite{StringOP}. Another example would be the transitions between SPT phases and their gauge theory duals. For example, the transition between the toric code and double semion ground states was studied in \cite{StripeOrder} and was shown to possess stripe order. Here, we expect $\langle{b_1}\rangle$ to be minimal at the center of the stripe order transition.

Finally, in the context of $3D$ Ising models, previous interest was shown in computing the entropy of surfaces (number of possible configurations possessing a certain genus or Betti number) \cite{Surfaces1,Surfaces2}. Following this homology computation under the VR complex, relative entropies of different values of the Betti number can be easily obtained. As an example, Figure \ref{fig:betti} provides a histogram showing the frequencies of various values of $b_1$ for Monte Carlo samples of the $\mathbb{Z}_2$ gauge theory with $D=3$ and $T=0.7$. Here, the temperature is selected such that no samples possess vison defects. 

We expect that the efficiency and ease of implementation of persistent homology will make it a useful tool in computational studies of a wide variety of topological systems in the future.  In addition, with the advent of highly controlled quantum devices which can deliver high-quality datasets of qubit measurement configurations \cite{Chen2021,Satzinger,RydbergSL}, we believe that persistent homology will be useful in identifying topological structures in present and near-term quantum computers exploring topological phases and phase transitions.

\section*{Acknowledgements}
We are grateful to J. McGreevy, T. Grover, Y.-Z. You, P. Fendley and E. Fradkin for critical discussions.  We also thank members of the Perimeter Institute Quantum Intelligence Lab for numerous insightful discussions.  This work was made possible by the facilities of the Shared Hierarchical Academic Research Computing Network (SHARCNET) and Compute Canada. This work was supported by NSERC, the Canada Research Chair program, and the Perimeter Institute for Theoretical Physics. Research at Perimeter Institute is supported in part by the Government of Canada through the Department of Innovation, Science and Economic Development Canada and by the Province of Ontario through the Ministry of Economic Development, Job Creation and Trade. 

\bibliography{apssamp.bib}

\begin{thebibliography}{31}%
\makeatletter
\providecommand \@ifxundefined [1]{%
 \@ifx{#1\undefined}
}%
\providecommand \@ifnum [1]{%
 \ifnum #1\expandafter \@firstoftwo
 \else \expandafter \@secondoftwo
 \fi
}%
\providecommand \@ifx [1]{%
 \ifx #1\expandafter \@firstoftwo
 \else \expandafter \@secondoftwo
 \fi
}%
\providecommand \natexlab [1]{#1}%
\providecommand \enquote  [1]{``#1''}%
\providecommand \bibnamefont  [1]{#1}%
\providecommand \bibfnamefont [1]{#1}%
\providecommand \citenamefont [1]{#1}%
\providecommand \href@noop [0]{\@secondoftwo}%
\providecommand \href [0]{\begingroup \@sanitize@url \@href}%
\providecommand \@href[1]{\@@startlink{#1}\@@href}%
\providecommand \@@href[1]{\endgroup#1\@@endlink}%
\providecommand \@sanitize@url [0]{\catcode `\\12\catcode `\$12\catcode
  `\&12\catcode `\#12\catcode `\^12\catcode `\_12\catcode `\%12\relax}%
\providecommand \@@startlink[1]{}%
\providecommand \@@endlink[0]{}%
\providecommand \url  [0]{\begingroup\@sanitize@url \@url }%
\providecommand \@url [1]{\endgroup\@href {#1}{\urlprefix }}%
\providecommand \urlprefix  [0]{URL }%
\providecommand \Eprint [0]{\href }%
\providecommand \doibase [0]{https://doi.org/}%
\providecommand \selectlanguage [0]{\@gobble}%
\providecommand \bibinfo  [0]{\@secondoftwo}%
\providecommand \bibfield  [0]{\@secondoftwo}%
\providecommand \translation [1]{[#1]}%
\providecommand \BibitemOpen [0]{}%
\providecommand \bibitemStop [0]{}%
\providecommand \bibitemNoStop [0]{.\EOS\space}%
\providecommand \EOS [0]{\spacefactor3000\relax}%
\providecommand \BibitemShut  [1]{\csname bibitem#1\endcsname}%
\let\auto@bib@innerbib\@empty
\bibitem [{\citenamefont {Levin}\ and\ \citenamefont {Wen}(2005)}]{Wen1}%
  \BibitemOpen
  \bibfield  {author} {\bibinfo {author} {\bibfnamefont {M.~A.}\ \bibnamefont
  {Levin}}\ and\ \bibinfo {author} {\bibfnamefont {X.-G.}\ \bibnamefont
  {Wen}},\ }\bibfield  {title} {\bibinfo {title} {String-net condensation: A
  physical mechanism for topological phases},\ }\href
  {https://doi.org/10.1103/PhysRevB.71.045110} {\bibfield  {journal} {\bibinfo
  {journal} {Phys. Rev. B}\ }\textbf {\bibinfo {volume} {71}},\ \bibinfo
  {pages} {045110} (\bibinfo {year} {2005})}\BibitemShut {NoStop}%
\bibitem [{\citenamefont {Wen}(2017)}]{Wen2}%
  \BibitemOpen
  \bibfield  {author} {\bibinfo {author} {\bibfnamefont {X.-G.}\ \bibnamefont
  {Wen}},\ }\bibfield  {title} {\bibinfo {title} {Colloquium: Zoo of
  quantum-topological phases of matter},\ }\href
  {https://doi.org/10.1103/RevModPhys.89.041004} {\bibfield  {journal}
  {\bibinfo  {journal} {Rev. Mod. Phys.}\ }\textbf {\bibinfo {volume} {89}},\
  \bibinfo {pages} {041004} (\bibinfo {year} {2017})}\BibitemShut {NoStop}%
\bibitem [{\citenamefont {Kitaev}(2003)}]{Kitaev}%
  \BibitemOpen
  \bibfield  {author} {\bibinfo {author} {\bibfnamefont {A.}~\bibnamefont
  {Kitaev}},\ }\bibfield  {title} {\bibinfo {title} {Fault-tolerant quantum
  computation by anyons},\ }\href
  {https://doi.org/https://doi.org/10.1016/S0003-4916(02)00018-0} {\bibfield
  {journal} {\bibinfo  {journal} {Annals of Physics}\ }\textbf {\bibinfo
  {volume} {303}},\ \bibinfo {pages} {2} (\bibinfo {year} {2003})}\BibitemShut
  {NoStop}%
\bibitem [{\citenamefont {Nayak}\ \emph {et~al.}(2008)\citenamefont {Nayak},
  \citenamefont {Simon}, \citenamefont {Stern}, \citenamefont {Freedman},\ and\
  \citenamefont {Das~Sarma}}]{Nayak}%
  \BibitemOpen
  \bibfield  {author} {\bibinfo {author} {\bibfnamefont {C.}~\bibnamefont
  {Nayak}}, \bibinfo {author} {\bibfnamefont {S.~H.}\ \bibnamefont {Simon}},
  \bibinfo {author} {\bibfnamefont {A.}~\bibnamefont {Stern}}, \bibinfo
  {author} {\bibfnamefont {M.}~\bibnamefont {Freedman}},\ and\ \bibinfo
  {author} {\bibfnamefont {S.}~\bibnamefont {Das~Sarma}},\ }\bibfield  {title}
  {\bibinfo {title} {Non-abelian anyons and topological quantum computation},\
  }\href {https://doi.org/10.1103/RevModPhys.80.1083} {\bibfield  {journal}
  {\bibinfo  {journal} {Rev. Mod. Phys.}\ }\textbf {\bibinfo {volume} {80}},\
  \bibinfo {pages} {1083} (\bibinfo {year} {2008})}\BibitemShut {NoStop}%
\bibitem [{\citenamefont {Balents}(2010)}]{Balents2010}%
  \BibitemOpen
  \bibfield  {author} {\bibinfo {author} {\bibfnamefont {L.}~\bibnamefont
  {Balents}},\ }\bibfield  {title} {\bibinfo {title} {Spin liquids in
  frustrated magnets},\ }\href {https://doi.org/10.1038/nature08917} {\bibfield
   {journal} {\bibinfo  {journal} {Nature}\ }\textbf {\bibinfo {volume}
  {464}},\ \bibinfo {pages} {199} (\bibinfo {year} {2010})}\BibitemShut
  {NoStop}%
\bibitem [{\citenamefont {Chen}\ \emph {et~al.}(2021)\citenamefont {Chen} \emph
  {et~al.}}]{Chen2021}%
  \BibitemOpen
  \bibfield  {author} {\bibinfo {author} {\bibfnamefont {Z.}~\bibnamefont
  {Chen}} \emph {et~al.},\ }\bibfield  {title} {\bibinfo {title} {Exponential
  suppression of bit or phase errors with cyclic error correction},\ }\href
  {https://doi.org/10.1038/s41586-021-03588-y} {\bibfield  {journal} {\bibinfo
  {journal} {Nature}\ }\textbf {\bibinfo {volume} {595}},\ \bibinfo {pages}
  {383} (\bibinfo {year} {2021})}\BibitemShut {NoStop}%
\bibitem [{\citenamefont {Satzinger}\ \emph {et~al.}(2021)\citenamefont
  {Satzinger} \emph {et~al.}}]{Satzinger}%
  \BibitemOpen
  \bibfield  {author} {\bibinfo {author} {\bibfnamefont {K.~J.}\ \bibnamefont
  {Satzinger}} \emph {et~al.},\ }\bibfield  {title} {\bibinfo {title}
  {Realizing topologically ordered states on a quantum processor},\ }\href
  {https://doi.org/10.1126/science.abi8378} {\bibfield  {journal} {\bibinfo
  {journal} {Science}\ }\textbf {\bibinfo {volume} {374}},\ \bibinfo {pages}
  {1237} (\bibinfo {year} {2021})}\BibitemShut {NoStop}%
\bibitem [{\citenamefont {Semeghini}\ \emph {et~al.}(2021)\citenamefont
  {Semeghini} \emph {et~al.}}]{RydbergSL}%
  \BibitemOpen
  \bibfield  {author} {\bibinfo {author} {\bibfnamefont {G.}~\bibnamefont
  {Semeghini}} \emph {et~al.},\ }\bibfield  {title} {\bibinfo {title} {Probing
  topological spin liquids on a programmable quantum simulator},\ }\href
  {https://doi.org/10.1126/science.abi8794} {\bibfield  {journal} {\bibinfo
  {journal} {Science}\ }\textbf {\bibinfo {volume} {374}},\ \bibinfo {pages}
  {1242} (\bibinfo {year} {2021})}\BibitemShut {NoStop}%
\bibitem [{\citenamefont {Levin}\ and\ \citenamefont {Wen}(2006)}]{TOPOEE1}%
  \BibitemOpen
  \bibfield  {author} {\bibinfo {author} {\bibfnamefont {M.}~\bibnamefont
  {Levin}}\ and\ \bibinfo {author} {\bibfnamefont {X.-G.}\ \bibnamefont
  {Wen}},\ }\bibfield  {title} {\bibinfo {title} {Detecting topological order
  in a ground state wave function},\ }\href
  {https://doi.org/10.1103/PhysRevLett.96.110405} {\bibfield  {journal}
  {\bibinfo  {journal} {Phys. Rev. Lett.}\ }\textbf {\bibinfo {volume} {96}},\
  \bibinfo {pages} {110405} (\bibinfo {year} {2006})}\BibitemShut {NoStop}%
\bibitem [{\citenamefont {Kitaev}\ and\ \citenamefont
  {Preskill}(2006)}]{TOPOEE2}%
  \BibitemOpen
  \bibfield  {author} {\bibinfo {author} {\bibfnamefont {A.}~\bibnamefont
  {Kitaev}}\ and\ \bibinfo {author} {\bibfnamefont {J.}~\bibnamefont
  {Preskill}},\ }\bibfield  {title} {\bibinfo {title} {Topological entanglement
  entropy},\ }\href {https://doi.org/10.1103/PhysRevLett.96.110404} {\bibfield
  {journal} {\bibinfo  {journal} {Phys. Rev. Lett.}\ }\textbf {\bibinfo
  {volume} {96}},\ \bibinfo {pages} {110404} (\bibinfo {year}
  {2006})}\BibitemShut {NoStop}%
\bibitem [{\citenamefont {Ghrist}(2008)}]{TDA1}%
  \BibitemOpen
  \bibfield  {author} {\bibinfo {author} {\bibfnamefont {R.}~\bibnamefont
  {Ghrist}},\ }\bibfield  {title} {\bibinfo {title} {Barcodes: The persistent
  topology of data},\ }\href
  {https://doi.org/https://doi.org/10.1090/S0273-0979-07-01191-3} {\bibfield
  {journal} {\bibinfo  {journal} {B. Am. Math. Soc.}\ }\textbf {\bibinfo
  {volume} {45}},\ \bibinfo {pages} {61} (\bibinfo {year} {2008})}\BibitemShut
  {NoStop}%
\bibitem [{\citenamefont {Zomorodian}\ and\ \citenamefont
  {Carlsson}(2005)}]{TDA2}%
  \BibitemOpen
  \bibfield  {author} {\bibinfo {author} {\bibfnamefont {A.}~\bibnamefont
  {Zomorodian}}\ and\ \bibinfo {author} {\bibfnamefont {G.}~\bibnamefont
  {Carlsson}},\ }\bibfield  {title} {\bibinfo {title} {Computing persistent
  homology},\ }\href {https://doi.org/10.1007/s00454-004-1146-y} {\bibfield
  {journal} {\bibinfo  {journal} {Discrete {\&} Computational Geometry}\
  }\textbf {\bibinfo {volume} {33}},\ \bibinfo {pages} {249} (\bibinfo {year}
  {2005})}\BibitemShut {NoStop}%
\bibitem [{\citenamefont {Carlsson}(2009)}]{TDA3}%
  \BibitemOpen
  \bibfield  {author} {\bibinfo {author} {\bibfnamefont {G.}~\bibnamefont
  {Carlsson}},\ }\bibfield  {title} {\bibinfo {title} {Topology and data},\
  }\href {https://doi.org/https://doi.org/10.1090/S0273-0979-09-01249-X}
  {\bibfield  {journal} {\bibinfo  {journal} {B. Am. Math. Soc.}\ }\textbf
  {\bibinfo {volume} {46}},\ \bibinfo {pages} {255} (\bibinfo {year}
  {2009})}\BibitemShut {NoStop}%
\bibitem [{\citenamefont {{Edelsbrunner}}\ \emph {et~al.}(2002)\citenamefont
  {{Edelsbrunner}}, \citenamefont {{Letscher}},\ and\ \citenamefont
  {{Zomorodian}}}]{TDA4}%
  \BibitemOpen
  \bibfield  {author} {\bibinfo {author} {\bibnamefont {{Edelsbrunner}}},
  \bibinfo {author} {\bibnamefont {{Letscher}}},\ and\ \bibinfo {author}
  {\bibnamefont {{Zomorodian}}},\ }\bibfield  {title} {\bibinfo {title}
  {Topological persistence and simplification},\ }\href
  {https://doi.org/10.1007/s00454-002-2885-2} {\bibfield  {journal} {\bibinfo
  {journal} {Discrete {\&} Computational Geometry}\ }\textbf {\bibinfo {volume}
  {28}},\ \bibinfo {pages} {511} (\bibinfo {year} {2002})}\BibitemShut
  {NoStop}%
\bibitem [{\citenamefont {Donato}\ \emph {et~al.}(2016)\citenamefont {Donato},
  \citenamefont {Gori}, \citenamefont {Pettini}, \citenamefont {Petri},
  \citenamefont {De~Nigris}, \citenamefont {Franzosi},\ and\ \citenamefont
  {Vaccarino}}]{PH1}%
  \BibitemOpen
  \bibfield  {author} {\bibinfo {author} {\bibfnamefont {I.}~\bibnamefont
  {Donato}}, \bibinfo {author} {\bibfnamefont {M.}~\bibnamefont {Gori}},
  \bibinfo {author} {\bibfnamefont {M.}~\bibnamefont {Pettini}}, \bibinfo
  {author} {\bibfnamefont {G.}~\bibnamefont {Petri}}, \bibinfo {author}
  {\bibfnamefont {S.}~\bibnamefont {De~Nigris}}, \bibinfo {author}
  {\bibfnamefont {R.}~\bibnamefont {Franzosi}},\ and\ \bibinfo {author}
  {\bibfnamefont {F.}~\bibnamefont {Vaccarino}},\ }\bibfield  {title} {\bibinfo
  {title} {Persistent homology analysis of phase transitions},\ }\href
  {https://doi.org/10.1103/PhysRevE.93.052138} {\bibfield  {journal} {\bibinfo
  {journal} {Phys. Rev. E}\ }\textbf {\bibinfo {volume} {93}},\ \bibinfo
  {pages} {052138} (\bibinfo {year} {2016})}\BibitemShut {NoStop}%
\bibitem [{\citenamefont {Sale}\ \emph {et~al.}(2021)\citenamefont {Sale},
  \citenamefont {Giansiracusa},\ and\ \citenamefont {Lucini}}]{Sale}%
  \BibitemOpen
  \bibfield  {author} {\bibinfo {author} {\bibfnamefont {N.}~\bibnamefont
  {Sale}}, \bibinfo {author} {\bibfnamefont {J.}~\bibnamefont {Giansiracusa}},\
  and\ \bibinfo {author} {\bibfnamefont {B.}~\bibnamefont {Lucini}},\
  }\bibfield  {title} {\bibinfo {title} {Quantitative analysis of phase
  transitions in two-dimensional xy models using persistent homology},\
  }\href@noop {} {\  (\bibinfo {year} {2021})},\ \Eprint
  {https://arxiv.org/abs/arXiv:2109.10960} {arXiv:2109.10960} \BibitemShut
  {NoStop}%
\bibitem [{\citenamefont {Tran}\ \emph {et~al.}(2021)\citenamefont {Tran},
  \citenamefont {Chen},\ and\ \citenamefont {Hasegawa}}]{PH2}%
  \BibitemOpen
  \bibfield  {author} {\bibinfo {author} {\bibfnamefont {Q.~H.}\ \bibnamefont
  {Tran}}, \bibinfo {author} {\bibfnamefont {M.}~\bibnamefont {Chen}},\ and\
  \bibinfo {author} {\bibfnamefont {Y.}~\bibnamefont {Hasegawa}},\ }\bibfield
  {title} {\bibinfo {title} {Topological persistence machine of phase
  transitions},\ }\href {https://doi.org/10.1103/PhysRevE.103.052127}
  {\bibfield  {journal} {\bibinfo  {journal} {Phys. Rev. E}\ }\textbf {\bibinfo
  {volume} {103}},\ \bibinfo {pages} {052127} (\bibinfo {year}
  {2021})}\BibitemShut {NoStop}%
\bibitem [{\citenamefont {Olsthoorn}\ \emph {et~al.}(2020)\citenamefont
  {Olsthoorn}, \citenamefont {Hellsvik},\ and\ \citenamefont {Balatsky}}]{PH3}%
  \BibitemOpen
  \bibfield  {author} {\bibinfo {author} {\bibfnamefont {B.}~\bibnamefont
  {Olsthoorn}}, \bibinfo {author} {\bibfnamefont {J.}~\bibnamefont
  {Hellsvik}},\ and\ \bibinfo {author} {\bibfnamefont {A.~V.}\ \bibnamefont
  {Balatsky}},\ }\bibfield  {title} {\bibinfo {title} {Finding hidden order in
  spin models with persistent homology},\ }\href
  {https://doi.org/10.1103/PhysRevResearch.2.043308} {\bibfield  {journal}
  {\bibinfo  {journal} {Phys. Rev. Research}\ }\textbf {\bibinfo {volume}
  {2}},\ \bibinfo {pages} {043308} (\bibinfo {year} {2020})}\BibitemShut
  {NoStop}%
\bibitem [{\citenamefont {Cole}\ \emph {et~al.}(2020)\citenamefont {Cole},
  \citenamefont {Loges},\ and\ \citenamefont {Shiu}}]{PH4}%
  \BibitemOpen
  \bibfield  {author} {\bibinfo {author} {\bibfnamefont {A.}~\bibnamefont
  {Cole}}, \bibinfo {author} {\bibfnamefont {G.~J.}\ \bibnamefont {Loges}},\
  and\ \bibinfo {author} {\bibfnamefont {G.}~\bibnamefont {Shiu}},\ }\href@noop
  {} {\bibinfo {title} {Quantitative and interpretable order parameters for
  phase transitions from persistent homology}} (\bibinfo {year} {2020}),\
  \Eprint {https://arxiv.org/abs/2009.14231} {arXiv:2009.14231
  [cond-mat.stat-mech]} \BibitemShut {NoStop}%
\bibitem [{\citenamefont {Kogut}(1979)}]{Kogut}%
  \BibitemOpen
  \bibfield  {author} {\bibinfo {author} {\bibfnamefont {J.~B.}\ \bibnamefont
  {Kogut}},\ }\bibfield  {title} {\bibinfo {title} {An introduction to lattice
  gauge theory and spin systems},\ }\href
  {https://doi.org/10.1103/RevModPhys.51.659} {\bibfield  {journal} {\bibinfo
  {journal} {Rev. Mod. Phys.}\ }\textbf {\bibinfo {volume} {51}},\ \bibinfo
  {pages} {659} (\bibinfo {year} {1979})}\BibitemShut {NoStop}%
\bibitem [{\citenamefont {Tupitsyn}\ \emph {et~al.}(2010)\citenamefont
  {Tupitsyn}, \citenamefont {Kitaev}, \citenamefont {Prokof'ev},\ and\
  \citenamefont {Stamp}}]{MCP}%
  \BibitemOpen
  \bibfield  {author} {\bibinfo {author} {\bibfnamefont {I.~S.}\ \bibnamefont
  {Tupitsyn}}, \bibinfo {author} {\bibfnamefont {A.}~\bibnamefont {Kitaev}},
  \bibinfo {author} {\bibfnamefont {N.~V.}\ \bibnamefont {Prokof'ev}},\ and\
  \bibinfo {author} {\bibfnamefont {P.~C.~E.}\ \bibnamefont {Stamp}},\
  }\bibfield  {title} {\bibinfo {title} {Topological multicritical point in the
  phase diagram of the toric code model and three-dimensional lattice gauge
  higgs model},\ }\href {https://doi.org/10.1103/PhysRevB.82.085114} {\bibfield
   {journal} {\bibinfo  {journal} {Phys. Rev. B}\ }\textbf {\bibinfo {volume}
  {82}},\ \bibinfo {pages} {085114} (\bibinfo {year} {2010})}\BibitemShut
  {NoStop}%
\bibitem [{\citenamefont {Adams}\ \emph {et~al.}(2017)\citenamefont {Adams},
  \citenamefont {Emerson}, \citenamefont {Kirby}, \citenamefont {Neville},
  \citenamefont {Peterson}, \citenamefont {Shipman}, \citenamefont
  {Chepushtanova}, \citenamefont {Hanson}, \citenamefont {Motta},\ and\
  \citenamefont {Ziegelmeier}}]{Stability}%
  \BibitemOpen
  \bibfield  {author} {\bibinfo {author} {\bibfnamefont {H.}~\bibnamefont
  {Adams}}, \bibinfo {author} {\bibfnamefont {T.}~\bibnamefont {Emerson}},
  \bibinfo {author} {\bibfnamefont {M.}~\bibnamefont {Kirby}}, \bibinfo
  {author} {\bibfnamefont {R.}~\bibnamefont {Neville}}, \bibinfo {author}
  {\bibfnamefont {C.}~\bibnamefont {Peterson}}, \bibinfo {author}
  {\bibfnamefont {P.}~\bibnamefont {Shipman}}, \bibinfo {author} {\bibfnamefont
  {S.}~\bibnamefont {Chepushtanova}}, \bibinfo {author} {\bibfnamefont
  {E.}~\bibnamefont {Hanson}}, \bibinfo {author} {\bibfnamefont
  {F.}~\bibnamefont {Motta}},\ and\ \bibinfo {author} {\bibfnamefont
  {L.}~\bibnamefont {Ziegelmeier}},\ }\bibfield  {title} {\bibinfo {title}
  {Persistence images: A stable vector representation of persistent homology},\
  }\href {http://jmlr.org/papers/v18/16-337.html} {\bibfield  {journal}
  {\bibinfo  {journal} {Journal of Machine Learning Research}\ }\textbf
  {\bibinfo {volume} {18}},\ \bibinfo {pages} {1} (\bibinfo {year}
  {2017})}\BibitemShut {NoStop}%
\bibitem [{\citenamefont {Tralie}\ \emph {et~al.}(2018)\citenamefont {Tralie},
  \citenamefont {Saul},\ and\ \citenamefont {Bar-On}}]{Ripser}%
  \BibitemOpen
  \bibfield  {author} {\bibinfo {author} {\bibfnamefont {C.}~\bibnamefont
  {Tralie}}, \bibinfo {author} {\bibfnamefont {N.}~\bibnamefont {Saul}},\ and\
  \bibinfo {author} {\bibfnamefont {R.}~\bibnamefont {Bar-On}},\ }\bibfield
  {title} {\bibinfo {title} {{Ripser.py}: A lean persistent homology library
  for python},\ }\href {https://doi.org/10.21105/joss.00925} {\bibfield
  {journal} {\bibinfo  {journal} {The Journal of Open Source Software}\
  }\textbf {\bibinfo {volume} {3}},\ \bibinfo {pages} {925} (\bibinfo {year}
  {2018})}\BibitemShut {NoStop}%
\bibitem [{\citenamefont {Cardy}(1996)}]{Cardy}%
  \BibitemOpen
  \bibfield  {author} {\bibinfo {author} {\bibfnamefont {J.}~\bibnamefont
  {Cardy}},\ }\href {https://doi.org/10.1017/CBO9781316036440} {\emph {\bibinfo
  {title} {Scaling and Renormalization in Statistical Physics}}},\ Cambridge
  Lecture Notes in Physics\ (\bibinfo  {publisher} {Cambridge University
  Press},\ \bibinfo {year} {1996})\BibitemShut {NoStop}%
\bibitem [{\citenamefont {Rodriguez-Nieva}\ and\ \citenamefont
  {Scheurer}(2019)}]{DM1}%
  \BibitemOpen
  \bibfield  {author} {\bibinfo {author} {\bibfnamefont {J.~F.}\ \bibnamefont
  {Rodriguez-Nieva}}\ and\ \bibinfo {author} {\bibfnamefont {M.~S.}\
  \bibnamefont {Scheurer}},\ }\bibfield  {title} {\bibinfo {title} {Identifying
  topological order through unsupervised machine learning},\ }\href@noop {}
  {\bibfield  {journal} {\bibinfo  {journal} {Nature Physics}\ }\textbf
  {\bibinfo {volume} {15}},\ \bibinfo {pages} {790} (\bibinfo {year}
  {2019})}\BibitemShut {NoStop}%
\bibitem [{\citenamefont {Somoza}\ \emph {et~al.}(2020)\citenamefont {Somoza},
  \citenamefont {Serna},\ and\ \citenamefont {Nahum}}]{Somoza}%
  \BibitemOpen
  \bibfield  {author} {\bibinfo {author} {\bibfnamefont {A.~M.}\ \bibnamefont
  {Somoza}}, \bibinfo {author} {\bibfnamefont {P.}~\bibnamefont {Serna}},\ and\
  \bibinfo {author} {\bibfnamefont {A.}~\bibnamefont {Nahum}},\ }\href@noop {}
  {\bibinfo {title} {Self-dual criticality in three-dimensional $\mathbb{Z}_2$
  gauge theory with matter}} (\bibinfo {year} {2020}),\ \Eprint
  {https://arxiv.org/abs/2012.15845} {arXiv:2012.15845 [cond-mat.stat-mech]}
  \BibitemShut {NoStop}%
\bibitem [{\citenamefont {Kennedy}\ and\ \citenamefont
  {Tasaki}(1992)}]{Tasaki}%
  \BibitemOpen
  \bibfield  {author} {\bibinfo {author} {\bibfnamefont {T.}~\bibnamefont
  {Kennedy}}\ and\ \bibinfo {author} {\bibfnamefont {H.}~\bibnamefont
  {Tasaki}},\ }\bibfield  {title} {\bibinfo {title} {Hidden
  ${\mathrm{z}}_{2}$\ifmmode\times\else\texttimes\fi{}${\mathrm{z}}_{2}$
  symmetry breaking in haldane-gap antiferromagnets},\ }\href
  {https://doi.org/10.1103/PhysRevB.45.304} {\bibfield  {journal} {\bibinfo
  {journal} {Phys. Rev. B}\ }\textbf {\bibinfo {volume} {45}},\ \bibinfo
  {pages} {304} (\bibinfo {year} {1992})}\BibitemShut {NoStop}%
\bibitem [{\citenamefont {den Nijs}\ and\ \citenamefont
  {Rommelse}(1989)}]{StringOP}%
  \BibitemOpen
  \bibfield  {author} {\bibinfo {author} {\bibfnamefont {M.}~\bibnamefont {den
  Nijs}}\ and\ \bibinfo {author} {\bibfnamefont {K.}~\bibnamefont {Rommelse}},\
  }\bibfield  {title} {\bibinfo {title} {Preroughening transitions in crystal
  surfaces and valence-bond phases in quantum spin chains},\ }\href
  {https://doi.org/10.1103/PhysRevB.40.4709} {\bibfield  {journal} {\bibinfo
  {journal} {Phys. Rev. B}\ }\textbf {\bibinfo {volume} {40}},\ \bibinfo
  {pages} {4709} (\bibinfo {year} {1989})}\BibitemShut {NoStop}%
\bibitem [{\citenamefont {Dupont}\ \emph {et~al.}(2021)\citenamefont {Dupont},
  \citenamefont {Gazit},\ and\ \citenamefont {Scaffidi}}]{StripeOrder}%
  \BibitemOpen
  \bibfield  {author} {\bibinfo {author} {\bibfnamefont {M.}~\bibnamefont
  {Dupont}}, \bibinfo {author} {\bibfnamefont {S.}~\bibnamefont {Gazit}},\ and\
  \bibinfo {author} {\bibfnamefont {T.}~\bibnamefont {Scaffidi}},\ }\bibfield
  {title} {\bibinfo {title} {Evidence for deconfined $u(1)$ gauge theory at the
  transition between toric code and double semion},\ }\href
  {https://doi.org/10.1103/PhysRevB.103.L140412} {\bibfield  {journal}
  {\bibinfo  {journal} {Phys. Rev. B}\ }\textbf {\bibinfo {volume} {103}},\
  \bibinfo {pages} {L140412} (\bibinfo {year} {2021})}\BibitemShut {NoStop}%
\bibitem [{\citenamefont {Dotsenko}\ \emph {et~al.}(1993)\citenamefont
  {Dotsenko}, \citenamefont {Windey}, \citenamefont {Harris}, \citenamefont
  {Marinari}, \citenamefont {Martinec},\ and\ \citenamefont
  {Picco}}]{Surfaces1}%
  \BibitemOpen
  \bibfield  {author} {\bibinfo {author} {\bibfnamefont {V.~S.}\ \bibnamefont
  {Dotsenko}}, \bibinfo {author} {\bibfnamefont {P.}~\bibnamefont {Windey}},
  \bibinfo {author} {\bibfnamefont {G.}~\bibnamefont {Harris}}, \bibinfo
  {author} {\bibfnamefont {E.}~\bibnamefont {Marinari}}, \bibinfo {author}
  {\bibfnamefont {E.}~\bibnamefont {Martinec}},\ and\ \bibinfo {author}
  {\bibfnamefont {M.}~\bibnamefont {Picco}},\ }\bibfield  {title} {\bibinfo
  {title} {Critical and topological properties of cluster boundaries in the 3d
  ising model},\ }\href {https://doi.org/10.1103/PhysRevLett.71.811} {\bibfield
   {journal} {\bibinfo  {journal} {Phys. Rev. Lett.}\ }\textbf {\bibinfo
  {volume} {71}},\ \bibinfo {pages} {811} (\bibinfo {year} {1993})}\BibitemShut
  {NoStop}%
\bibitem [{\citenamefont {Dotsenko}\ \emph {et~al.}(1994)\citenamefont
  {Dotsenko}, \citenamefont {Harris}, \citenamefont {Marinari}, \citenamefont
  {Martinec}, \citenamefont {Picco},\ and\ \citenamefont {Windey}}]{Surfaces2}%
  \BibitemOpen
  \bibfield  {author} {\bibinfo {author} {\bibfnamefont {V.~S.}\ \bibnamefont
  {Dotsenko}}, \bibinfo {author} {\bibfnamefont {G.}~\bibnamefont {Harris}},
  \bibinfo {author} {\bibfnamefont {E.}~\bibnamefont {Marinari}}, \bibinfo
  {author} {\bibfnamefont {E.}~\bibnamefont {Martinec}}, \bibinfo {author}
  {\bibfnamefont {M.}~\bibnamefont {Picco}},\ and\ \bibinfo {author}
  {\bibfnamefont {P.}~\bibnamefont {Windey}},\ }\href@noop {} {\bibinfo {title}
  {The phenomenology of strings and clusters in the 3-d ising model}} (\bibinfo
  {year} {1994}),\ \Eprint {https://arxiv.org/abs/arXiv:hep-th/9401129}
  {arXiv:hep-th/9401129} \BibitemShut {NoStop}%
\end{thebibliography}%

\appendix

\section{\label{app:topology}Review of Topology}

\subsection{\label{app:homotopy}Homotopy}

\begin{figure}
    \centering
    \includegraphics[width=0.45\textwidth]{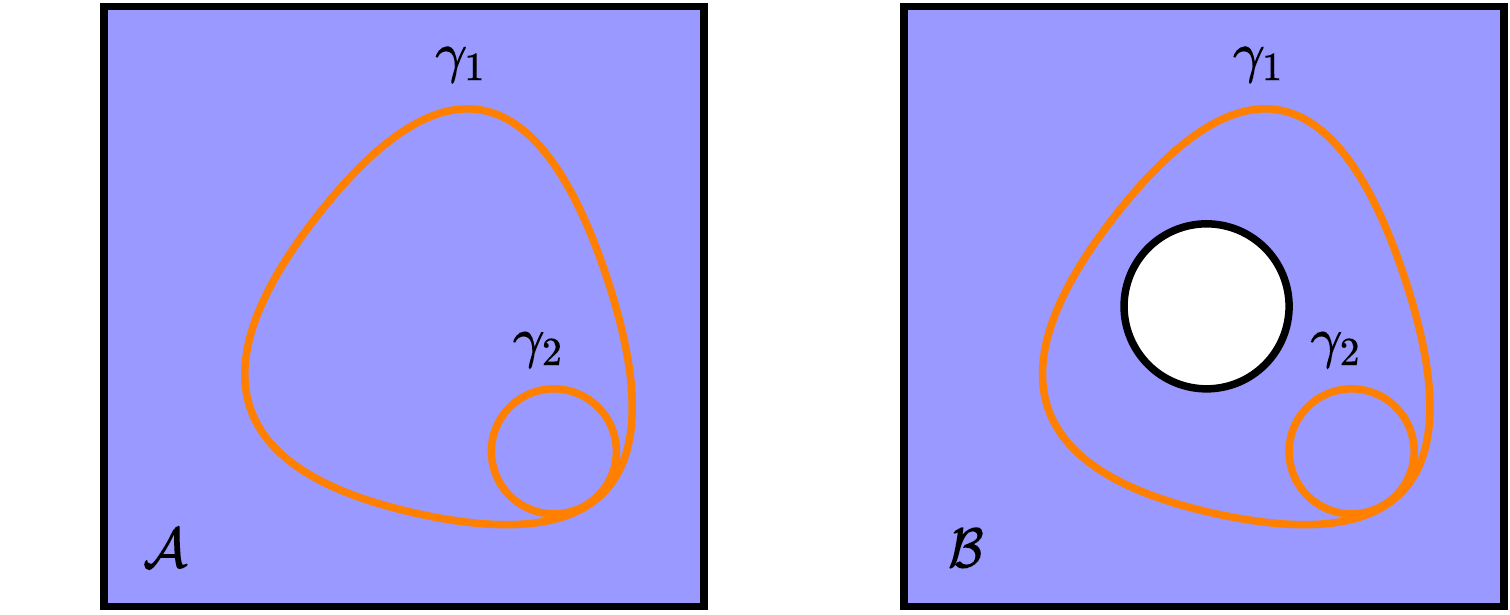}
    \caption{Two manifolds $\mathcal{A}$ and $\mathcal{B}$ and two curves $\gamma_1$ and $\gamma_2$. In $\mathcal{A}$, $\gamma_1$ and $\gamma_2$ are homotopically equivalent, since they can be smoothly deformed into each other. However, in $\mathcal{B}$, $\gamma_1$ and $\gamma_2$ are not homotopically equivalent, since $\gamma_2$ winds around a hole while $\gamma_1$ does not. Generally, there exists a homotopy class for each winding number around any given hole.}
    \label{fig:homotopy}
\end{figure}

Homotopy provides a rigorous formalism for establishing topological equivalence between objects, and is based on the notion of equivalence classes of loops. We say that any two loops $\gamma_1(t)$ and $\gamma_2(t)$ defined on a manifold $\mathcal{M}$ belong to the same \textit{homotopy class} if there exists a continuous function $H(t,s)$ on $\mathcal{M}$ such that $H(t,0)=\gamma_1(t)$ and $H(t,1)=\gamma_2(t)$. As examples, we consider the two manifolds shown in Figure \ref{fig:homotopy}, where $\mathcal{B}$ differs from $\mathcal{A}$ by the presence of a hole. One can draw infinitely many loops on $\mathcal{A}$, but all such loops can be continuously deformed into each other. Since all possible loops on $\mathcal{A}$ are contractible to a single point, their respective homotopy class is labelled the \textit{trivial class}.

Now consider $\mathcal{B}$. Any loop that encloses the hole is not contractible to a single point, and hence belongs to a nontrivial class. In fact, there exists a new nontrivial class for each winding number around the hole. The set of all homotopy classes for a given manifold can be used to form a group, where the trivial class is the identity element and the group operation is the addition of loops. For example, adding two loops of winding numbers $m$ and $n$ forms a loop of winding number $m+n$. Such a group of a manifold $\mathcal{M}$ is called the \textit{homotopy group} of $\mathcal{M}$ or $\pi_1(\mathcal{M})$. Since the homotopy group of $\mathcal{A}$ consists solely of the trivial class, $\pi_1(\mathcal{A})\cong{0}$. Since the homotopy group of $\mathcal{B}$ consists of an additional homotopy class for each winding number $n\in\mathbb{Z}$, $\pi_1(\mathcal{B})\cong\mathbb{Z}$.

The subscript in $\pi_1$ refers to the fact that 1-dimensional loops are used to construct homotopy classes. $\pi_1$ is called the first homotopy group or the fundamental group. It is possible to construct higher order homotopy groups from $n$-loops, these are denoted $\pi_n$. Higher order homotopy groups are necessary for detecting higher order defects. For example, consider the circle $S^1$ and the sphere $S^2$. 1-loops are capable of detecting the hole in $S^1$, and ultimately $\pi_1(S^1)\cong\mathbb{Z}$ following the same line of reasoning as above. However, 1-loops are not capable of detecting the higher dimensional hole in $S^2$: all 1-loops are contractible to a single point. Hence, $\pi_1(S^2)\cong{0}$. Instead, one can use 2-loops (two dimensional closed surfaces). In analogy with the discussion above, there exists a homotopy class of 2-loops for each winding number around the 3-dimensional hole. Hence, $\pi_2(S^2)\cong\mathbb{Z}$. 

Generally, a manifold $\mathcal{M}$ with $m$ holes of dimension $n$ has a homotopy group $\pi_{n-1}(\mathcal{M})=\mathbb{Z}\oplus\cdots\oplus\mathbb{Z}\equiv\mathbb{Z}^m$, since a set of nontrivial winding classes exists for each hole. Ultimately, these homotopy groups allow for a rigorous notion of topological equivalence. Namely, two objects $\mathcal{A}$ and $\mathcal{B}$ are topologically equivalent if $\pi_n(\mathcal{A})\cong\pi_n(\mathcal{B})$ for all $n$.

\subsection{\label{app:homology}Homology}

\begin{figure}
    \centering
    \includegraphics[width=0.25\textwidth]{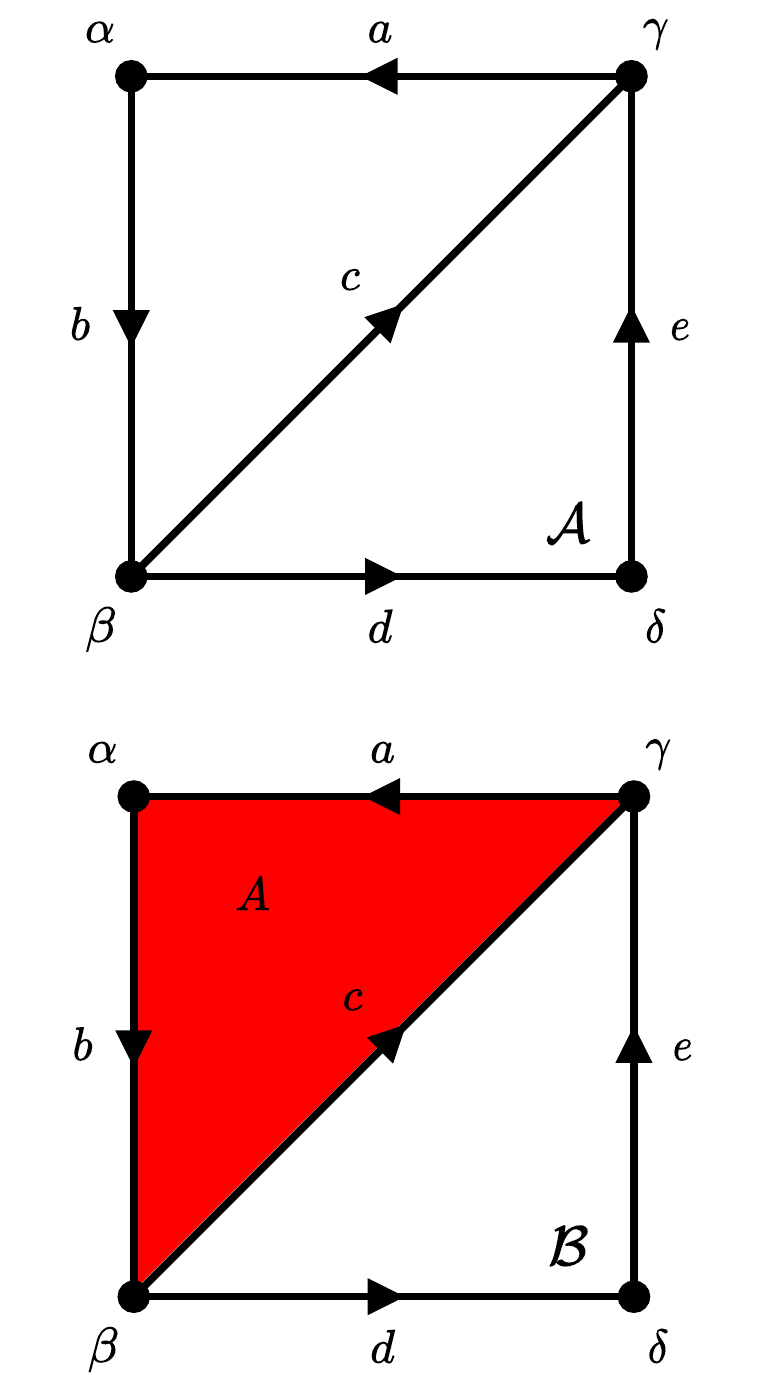}
    \caption{Two graphs $\mathcal{A}$ and $\mathcal{B}$. 0-cells (points) and 1-cells (lines) are labelled by greek and roman letters respectively. $\mathcal{B}$ differs from $\mathcal{A}$ by the presence of an additional 2-cell, labelled $A$, which fills one of the cycles.}
    \label{fig:homology}
\end{figure}

Homology is a framework for computing homotopy. To begin with, let us consider a general graph $X$ constructed from cells. Generally, an $n$-cell is an $n$-dimensional object, and the boundaries of an $n$-cell are $n-1$ cells. For example, a 0-cell is a point, a 1-cell is a directed line, and a 2-cell is a oriented disk. Any combination of $n$-cells is then referred to as $n$-chain of the graph $X$. If $C_n$ denotes the set of all $n$-chains of a graph $X$, and $\partial_n$ is the boundary operator of $n$-chains, it follows that $\partial_n:C_n\rightarrow{C_{n-1}}$. The $n^\text{th}$ homotopy group of the graph $X$ can then be computed via the $n^\text{th}$ homology group of $X$:
\begin{equation}
        H_n(X)=\frac{Z_n(X)}{B_n(X)}=\frac{\ker\left ( \partial_n \right )}{\text{im}\left ( \partial_{n+1} \right )}
\end{equation}
where $Z_n(X)=\ker\left ( \partial_n \right )$ refers to the $n$-dimensional cycles of $X$ (called $n$-cycles) and $B_n(X)=\text{im}\left ( \partial_{n+1} \right )$ refers to $n$-dimensional boundaries of $X$ (called $n$-boundaries). To understand the motivation behind this, we consider the two graphs shown in Figure \ref{fig:homology}. Both graphs possess the same 0-chains and 1-chains, where 0-cells and 1-cells are labelled by greek and roman letters respectively. However, graph $\mathcal{B}$ possesses an additional 2-cell labelled by $A$. The boundaries of each of the 1-cells are as follows:
\begin{align*}
\partial_1(a)&=\alpha-\gamma \\
\partial_1(b)&=\beta-\alpha \\
\partial_1(c)&=\gamma-\beta \\
\partial_1(d)&=\delta-\beta \\
\partial_1(e)&=\gamma-\delta
\end{align*} 
Any 1-cycle satisfies $\partial_1=0$. For example, $a+b+c$ forms a 1-cycle, since $\partial_1(a+b+c)=0$. To determine all 1-cycles, one can then solve $\partial_1(n_1a+n_2b+n_3c+n_4d+n_5e)=0$ where $n_i\in\mathbb{Z}$, which yields a system of linear equations. This gives $\ker\partial_1=\left \langle a+b+c,a+b+d+e \right \rangle\cong{\mathbb{Z}\oplus\mathbb{Z}}\equiv\mathbb{Z}^2$, implying their are 2 independent cycles. Since there are no 2-chains, $B_1(\mathcal{A})=0$. Hence, the homology of $\mathcal{A}$ is given by:
\begin{equation}
        H_1(\mathcal{A})=\frac{Z_1(\mathcal{A})}{B_1(\mathcal{A})}\cong\frac{\mathbb{Z}^2}{0}=\mathbb{Z}^2
\end{equation}
This is the result one would expect for homotopy: there exists a set of nontrivial homotopy classes for each independent cycle, since each independent cycle corresponds to a hole. However, the process has now been reduced to one of simple linear algebra.

Now let us consider $\mathcal{B}$. Since $\mathcal{A}$ and $\mathcal{B}$ are equivalent in terms of 0-chains and 1-chains, $Z_1(\mathcal{A})=Z_1(\mathcal{B})$. However, $\mathcal{B}$ possesses an additional 2-chain given by $A$. Hence, $B_1(\mathcal{B})$ is no longer zero. Instead, since $\partial_2(A)=a+b+c$, it follows that $B_1(\mathcal{B})=\text{im}\partial_2=\left \langle a+b+c \right \rangle\cong\mathbb{Z}$. Hence, the homology of $\mathcal{B}$ is given by:
\begin{equation}
        H_1(\mathcal{B})=\frac{Z_1(\mathcal{B})}{B_1(\mathcal{B})}\cong\frac{\mathbb{Z}^2}{\mathbb{Z}}=\mathbb{Z}
\end{equation}
Once again, this is precisely the expected result for homotopy. Namely, only one of the independent cycles contributes a set of nontrivial winding classes, since the other is filled by a 2-cell. This is the motivation behind computing cycles mod boundaries. Any independent cycle is a hole and hence contributes a set of nontrivial winding classes, unless it is filled by a higher dimensional cell.

To generalize this method to any smooth manifold, one can apply a preliminary \textit{triangulation} step, in which the the manifold is first transformed into a simplicial complex using a homeomorphism (see Section \ref{app:simplices}). This general procedure of applying a homology computation following a triangulation is referred to as \textit{simplicial homology}.

\subsection{\label{app:simplices}Simplices}

\begin{figure}
    \centering
    \includegraphics[width=0.35\textwidth]{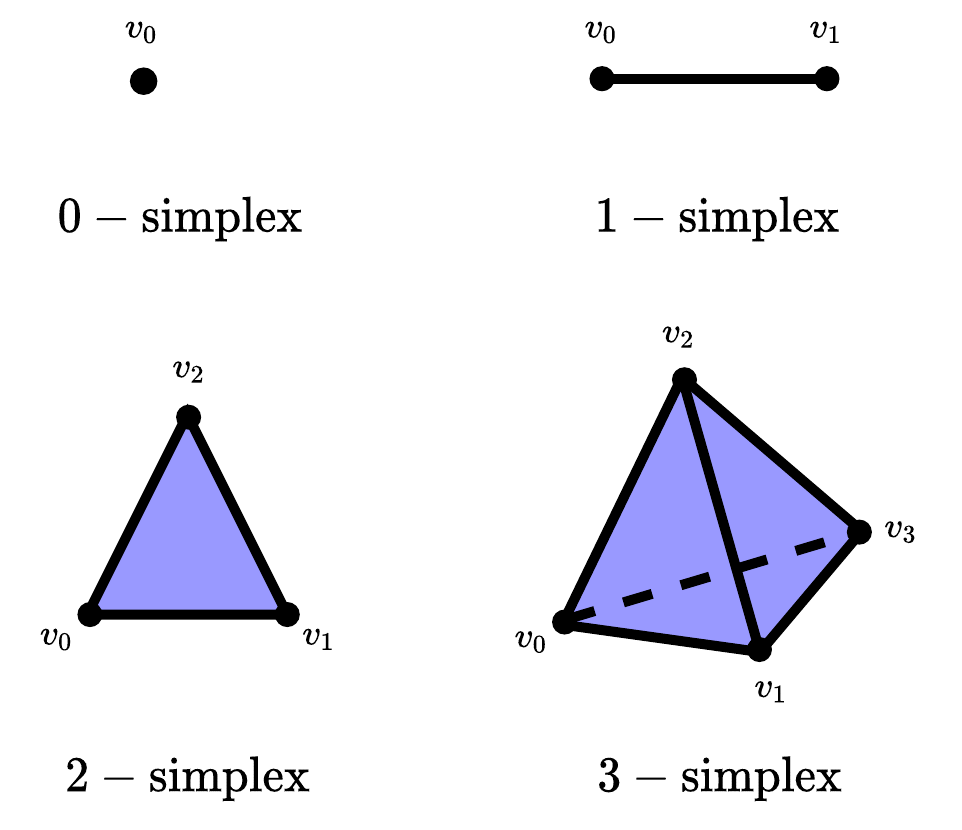}
    \caption{Illustration of first four simplices. The faces of any $n$-simplex are $n-1$ simplices. Both the 2-simplex and the 3-simplex are completely filled.}
    \label{fig:simplices}
\end{figure}

A simplex is a higher dimension generalization of a triangle. If $\left \{ \bm{v}_0,...,\bm{v}_n \right \}$ is a set of $n+1$ points or vertices, then an $n$-simplex $\Delta_n$ is described by the convex hull of these points: $\sum_ic_i\bm{v}_i$ where $\sum_ic_i=1$ and $0\leq{c_i}\leq{1}$ are the \textit{barycentric coordinates}. Following this definition, a 0-simplex is a point, a 1-simplex is a line, a 2-simplex is a filled triangle, and a 3-simplex is a filled tetrahedron (see Figure \ref{fig:simplices}). It additionally follows that the faces of any $n$-simplex are $(n-1)$-simplices. More concretely, the boundary of any $n$-simplex is given by:
\begin{equation}
    \partial(\Delta_n)=\sum_{\alpha=0}^n(-1)^\alpha\left ( v_0\cdots\hat{v}_i\cdots{v_n} \right )
\end{equation}
where the hat symbol denotes omit and each term in the sum denotes an oriented face of the simplex. Following this definition, a \textit{simplicial complex} is simply a set of connected and disconnected simplices satisfying the following rule: any two simplices that intersect must share a common face.

\end{document}